# Dyadic Green's function for a topological insulator stratified sphere

Huai-Yi Xie*

Division of Physics, Institute of Nuclear Energy Research, Taoyuan County 32546, Taiwan

*Corresponding author: damoxie156@gmail.com

(Date: 2023/9/12)

**Abstract**

We construct the dyadic Green's functions (DGFs) for a topological insulator (TI) stratified sphere within the framework of axion electrodynamics. For these DGFs, the additional expansion coefficients are included to account for the axion coupling effect. With the application of these DGFs, we derive the formulation of light scattering from a dipole near a TI stratified sphere. In our numerical studies, we give three types of configurations (a metal-coated TI sphere, a metal-TI-metal-coated TI sphere and an alternating metal-TI stratified sphere) to investigate how the topological magneto-electric (TME) response of the TI sphere (shells) influences on the multipolar plasmonic resonance of the metal shells. For these types, the results show that the TME effect causes some modifications of the decay rate spectrum for an emitting dipole near a TI stratified sphere. For the multipolar resonances of the metal shells, it is observed that the TME-induced red-shifts for the bonding and lower order antibonding modes are found but those for the higher order antibonding modes are insignificant. In addition, for a metal-coated TI sphere, we take into account the effects of losses in the TI core of which the dielectric function is chosen to be the form of the bulk or five quintuple layers (5QL) slab and then the some modifications of the TME-induced decay rate spectrum are obviously suppressed. These phenomenological characteristics provide useful guidance to probing the TME effect via molecular fluorescence experiments.

**Introduction**

The hypothetical elementary axion particle was first proposed about forty years ago to account for the strong CP problem in quantum chromodynamics (QCD) [1-3] and is regarded as a candidate for dark matter in cosmology [4-8]. Sikivie first formulated the axion electrodynamics (AED) including axion-photon interactions and proposed possible electromagnetic experiments to detect this elusive particle [9]. Subsequently, Wilczek discussed a possible application of AED to understand the unusual antiphase boundary in PbTe, and to extend its relevance to condensed matter physics via the quantized Hall effect [10]. However, it was not until about fifteen years ago with the discovery of topological insulators that many intriguing results within the framework of AED came to the spotlight.

Topological insulator (TI) is a new quantum state of (2D/3D) matter with an insulating interior but conducting surface states protected by time reversal (TR) symmetry [11-16], as manifested in composite materials of heavy atoms such as $Bi_2Te_3$, $Bi_2Se_3$, and $Sb_2Te_3$ [17-18]. When adding a weak magnetic field to the TI, doping the TI with magnetic impurities or coating the TI with a thin magnetic layer, we observe the topological magneto-electric (TME) effect due to the TR symmetry broken [13,19-22]. This effect leads to some unusual optical properties such as Faraday and Kerr polarization rotations at the Terahertz frequency, as observed in many intriguing experiments [23-25]. Moreover, the TME-modified scattering properties from a TI target with the far-field sources have been widely studied in the literature [20,26-28]. The TME-induced strong backward scattering is found in both Rayleigh and Mie scattering regimes for a TI cylinder [20]/sphere [26]. However, the TME effect is so small since it is associated with the fine structure constant. With the aid of plasmonic materials for the coreshell structure such as a metal-coated TI sphere [27] and TI-coated metal cylinder [28], the TME effect is enhanced by plasmonic resonances at the surface of the metal. For these structures, the observable TME-induced red-shifts are manifested in the scattering spectrum. In addition, with the dipole source near a metal-coated TI sphere, the TME effect causes red-shifts in the fluorescence emission spectrum significantly for the bonding modes of the metal shell with the lower order multipolar resonances [21].

As is well-known, Green's function method is a powerful tool for obtaining the field produced by a point source [29]. With the application of AED, the electrostatics Green's functions for different geometries of a TI target are formulated [30-31]. Optical intersubband properties of a semiconductor-coated TI sphere are studied via these Green's functions [32]. Including the retardation effect of electromagnetic waves, the dyadic Green's functions (DGFs) for planar TI multilayer films are constructed, with their coefficients being in connection to the modified Fresnel coefficients [33] as well as their application to studying the spontaneous emission of an emitting dipole near the surface of a two-level system [34] and planar TI multilayers [35]. However, to reveal TME effects, we go beyond the above planar TI multilayers to furthermore focus on TI nanoparticles systems [36] such as a TI stratified sphere. As mentioned above, the TME effect is enhanced by localized surface plasmonic resonances at the surface of the metal shells [37-38]. Since localized surface plasmons will enhance the electromagnetic fields in the near zone, there are many practical applications in the aspects of measurement in near field optics such as surface enhanced Raman spectroscope [39-40] and it is possible to open a channel to study the measurable TME-modified modification for plasmonic modes [21,26-27]. Moreover, it is well-known that a stratified sphere is one of the most versatile systems in its tunability of the plasmonic resonance modes [41]. Hence, we should develop an accurate and efficient method based on the DGFs for a TI stratified sphere to investigate how the TME-induced modification of these modes. However, to our knowledge, the DGFs for a TI stratified sphere have not been found in the literature.

For this reason, it is the purpose of our present work to study how to construct the DGFs for a TI stratified sphere, and we use them to formulate light scattering from a dipole near this structure. With the application of these DGFs, we first extend our previous work [21] to the study of the TME effect on the decay rate of an emitting dipole near a metal-coated TI sphere, including the retardation effect. To go beyond two material regions as presented in Ref. [21], we give more complicated configurations such as a metal-TI-metal-coated TI sphere and an alternating metal-TI stratified sphere. Next, for a metal-coated TI sphere, we consider the effects of losses in the TI core of which the dielectric function is chosen to be the form of the bulk [42] or five quintuple layers (5QL) slab [42] to investigate how the TME effect affects such a coreshell system.

**Constructing electric DGFs of a TI stratified sphere**

We begin by recalling how to derive the appropriate boundary conditions in the presence of TI within the framework of AED and use them to construct electric DGFs of a TI stratified sphere. Next, we use the transfer matrix method to solve all unknown expansion coefficients in these electric DGFs.

*(i) Appropriate boundary conditions*

Within the framework of axion electrodynamics [13,43-44], the total Lagrangian density $\mathcal{L}$ is represented as $\mathcal{L} = \mathcal{L}_0 + \mathcal{L}_{axion}$ where $\mathcal{L}_0$ is the conventional Lagrangian density and $\mathcal{L}_{axion}$ is the pseudo scalar product term coupling both the electric and magnetic fields as follows:

$$\mathcal{L}_{axion} = \frac{\alpha\Theta(\mathbf{r})}{4\pi^2}\mathbf{E}(\mathbf{r},t)\bullet\mathbf{B}(\mathbf{r},t), \qquad (1)$$

where $\alpha$ is the fine structure constant and $\Theta(\mathbf{r})$ is the axion parameter, which is taken to be time-independent since the characteristic frequency of TI is in the THz regime. Hence the modified Maxwell's equations for fields with harmonic time dependence ($e^{-i\omega t}$) as applied to TI are expressed in the following form (in Gaussian units) [21]:

$$\begin{aligned}
\nabla\times\mathbf{E} - \frac{i\omega}{c}\mathbf{B} &= 0 \\
\nabla\times\mathbf{H} + \frac{i\omega}{c}\mathbf{D} &= \frac{4\pi}{c}\mathbf{J} + \frac{\alpha}{\pi}\nabla\Theta\times\mathbf{E}, \\
\nabla\bullet\mathbf{B} &= 0 \\
\nabla\bullet\mathbf{D} &= 4\pi\rho - \frac{\alpha}{\pi}\nabla\Theta\bullet\mathbf{B}
\end{aligned} \qquad (2)$$

with the conventional constituent relations $\mathbf{D} = \varepsilon(\mathbf{r})\mathbf{E}$ and $\mathbf{H} = \mathbf{B}$, describing all media including a TI material to be linear and nonmagnetic. The alternative descriptions of the electrodynamics for TI are to use the conventional Maxwell equations with the modified constituent relations $\mathbf{D} = \varepsilon(\mathbf{r})\mathbf{E} + \frac{\alpha\Theta}{\pi}\mathbf{B}$ and $\mathbf{H} = \mathbf{B} - \frac{\alpha\Theta}{\pi}\mathbf{E}$ [28,33-35]. Note that the $P_3$ field in Ref. [13] is related to our axion field via $P_3 = -\frac{\Theta}{2\pi}$.

To obtain the appropriate boundary conditions for the fields in the presence of TI, without loss of generality, we consider the geometry consisting of two regions $\Omega_1$ and $\Omega_2$ with the boundary $S = \Omega_1 \cap \Omega_2$ on which $\mathbf{n}$ is a normal unit vector, as shown in Fig. 1. The axion parameter and the dielectric function in $\Omega_i, i = 1,2$ are taken to be the constant values $\Theta_i, i = 1,2$ and $\varepsilon_i, i = 1,2$, respectively. In a similar way as in deriving the boundary conditions for the fields from the conventional Maxwell's equations without source at the boundary $S$, Eq. (2) leads to the following appropriate boundary conditions for TI [21,27]:

$$\mathbf{n} \bullet (\varepsilon_2 \mathbf{E}_2 - \varepsilon_1 \mathbf{E}_1) = -\frac{\alpha}{\pi} \mathbf{n} \bullet \mathbf{B}(\Theta_2 - \Theta_1)$$

$$\mathbf{n} \bullet \mathbf{B}_2 = \mathbf{n} \bullet \mathbf{B}_1 = \mathbf{n} \bullet \mathbf{B}$$

$$\mathbf{n} \times \mathbf{E}_2 = \mathbf{n} \times \mathbf{E}_1 = \mathbf{n} \times \mathbf{E}$$

$$\mathbf{n} \times (\mathbf{B}_2 - \mathbf{B}_1) = \frac{\alpha}{\pi} \mathbf{n} \times \mathbf{E}(\Theta_2 - \Theta_1)$$

, (3)

where the subscript 1(2) is related to region 1(2). $-\frac{\alpha}{\pi} \mathbf{n} \bullet \mathbf{B}(\Theta_2 - \Theta_1)$ and $\frac{\alpha}{\pi} \mathbf{n} \times \mathbf{E}(\Theta_2 - \Theta_1)$ are the induced quantum Hall charge density $\rho_{ind}$ and current $\mathbf{J}_{ind}$ at the surface $S$, satisfying the continuity equation $\nabla \bullet \mathbf{J}_{ind} = i\omega \rho_{ind}$. If $\Theta_1 = \Theta_2$, it is obvious that Eqs. (2) and (3) reduce to the conventional Maxwell equations with the corresponding boundary conditions as well as $\rho_{ind} = 0$ and $\mathbf{J}_{ind} = 0$, reconfirming the fact that a homogeneous axion field leads to no observable physical effect since the pseudo scalar product term $\mathcal{L}_{axion}$ simply reduces to a total differential in this case [13,45].

*(ii) DGF formulation*

We consider a TI stratified sphere composed of $N$ concentric layers from $i = 1$ for a core to $i = N$ for an outer layer as shown in Fig. 2. The $i$-th layer has an outer radius $R_i$. The material of this $i$-th layer is assumed to be a homogeneous, isotropic and nonmagnetic medium, characterized by the constant dielectric function $\varepsilon_i$ with the wavenumber $k_i = \frac{\omega}{c}\sqrt{\varepsilon_i}$ and the constant axion parameter $\Theta_i$. The host medium is described by the dielectric constant $\varepsilon_{N+1}$ and $\Theta_{N+1} = 0$. Note that $\Theta = 0$ represents a traditional material. Next we represent the electric DGF as $\mathbf{G}_e^{(f,s)}(\mathbf{r},\mathbf{r}')$, putting the source $s$ (with the position vector $\mathbf{r}'$) into the outer region $s = N+1$ and the field $f$ (with the position vector $\mathbf{r}$) is in an arbitrary region ($f = 1,2,...,N,N+1$). According to the scattering superposition method [46-48], the electric DGF $\mathbf{G}_e^{(f,N+1)}$ can be written as the sum of the free-space electric DGF $\mathbf{G}_{e0}$ and the scattering electric DGF $\mathbf{G}_{es}^{(f,N+1)}$ in the following form:

$$\mathbf{G}_e^{(f,N+1)}(\mathbf{r},\mathbf{r}') = \mathbf{G}_{e0}(\mathbf{r},\mathbf{r}')\delta_{f,N+1} + \mathbf{G}_{es}^{(f,N+1)}(\mathbf{r},\mathbf{r}'), \quad (4)$$

where $\delta_{p,N+1}$ denotes the Kronecker delta function with the property

$\delta_{p,N+1} = \begin{cases} 1, p = N+1 \\ 0, p \neq N+1 \end{cases}$. To obtain the explicit forms of $\mathbf{G}_{e0}$ and $\mathbf{G}_{es}^{(f,N+1)}$ in the spherical coordinates, both transverse spherical vector wave functions $\mathbf{M}$ and $\mathbf{N}$ are adopted as follows [49]:

$$\begin{aligned} \mathbf{M}_{\sigma\ell m}^{(t)}(k_f) &= \nabla \times \left(\psi_{\sigma\ell m}^{(t)} \mathbf{r}\right) \\ \mathbf{N}_{\sigma\ell m}^{(t)}(k_f) &= \frac{1}{k_f} \nabla \times \nabla \times \left(\psi_{\sigma\ell m}^{(t)} \mathbf{r}\right) \end{aligned} \quad (5)$$

where $\mathbf{r}$ is the position vector. $\psi$ is the scalar wave function satisfying the scalar wave equation, and its explicit form in spherical coordinates $(r,\theta,\varphi)$ is:

$$\psi_{\sigma\ell m}^{(t)} = z_\ell^{(t)}(k_f r) P_\ell^m(\cos\theta) \begin{cases} \cos m\varphi \\ \sin m\varphi \end{cases}. \quad (6)$$

Here the subscript $\sigma$ is denoted by $e$ (even) or $o$ (odd), according to whether $\cos m\varphi$ or $\sin m\varphi$ is used. $z_\ell^{(t)}$ is either the Bessel function $j_\ell$ ($t=1$) or the first kind of the Hankel function $h_\ell^{(1)}$ ($t=3$). $P_\ell^m$ is the associated Legendre function. Note that the complete expressions of Eq. (5) are shown in Appendix A.

Since the free-space electric DGF $\mathbf{G}_{e0}$ satisfies just the conventional wave equation for traditional materials, we can use the method of $\mathbf{G}_m$ (the detail deviations are shown in Appendix B) to obtain the following expansion [46-48]:

$$\begin{aligned} \mathbf{G}_{e0}(\mathbf{r},\mathbf{r}') = &-\frac{4\pi}{ck_{N+1}^2}\delta(\mathbf{r}-\mathbf{r}')\mathbf{e}_r\mathbf{e}_r \\ &+\frac{ik_{N+1}}{c}\sum_{\sigma\ell m}\tilde{C}_{\ell m}\begin{cases} \left[\mathbf{M}_{\sigma\ell m}^{(1)}(k_{N+1})\mathbf{M}_{\sigma\ell m}^{\prime(3)}(k_{N+1}) + \mathbf{N}_{\sigma\ell m}^{(1)}(k_{N+1})\mathbf{N}_{\sigma\ell m}^{\prime(3)}(k_{N+1})\right], r<r' \\ \left[\mathbf{M}_{\sigma\ell m}^{(3)}(k_{N+1})\mathbf{M}_{\sigma\ell m}^{\prime(1)}(k_{N+1}) + \mathbf{N}_{\sigma\ell m}^{(3)}(k_{N+1})\mathbf{N}_{\sigma\ell m}^{\prime(1)}(k_{N+1})\right], r>r' \end{cases} \end{aligned} \quad (7)$$

where the summation symbol $\sum_{\sigma\ell m}$ denotes $\sum_{\sigma=e,o}\sum_{\ell=1}^{\infty}\sum_{m=0}^{\ell}$, the superscript $'$ represents the source coordinates, and the constant $\tilde{C}_{\ell m}$ is:

$$\tilde{C}_{\ell m} = (2-\delta_0)\frac{2\ell+1}{\ell(\ell+1)}\frac{(\ell-m)!}{(\ell+m)!}, \quad (8)$$

with $\delta_0 = \begin{cases} 1, m=0 \\ 0, m\neq 0 \end{cases}$. Moreover, referring to the mathematical formulation as shown in Eq. (7), the scattering electric DGF $\mathbf{G}_{es}^{(f,N+1)}$ in each region $f$ can be written in

the following forms:

$$\mathbf{G}_{es}^{(N+1,N+1)}(\mathbf{r},\mathbf{r}') = \frac{ik_{N+1}}{c}\sum_{\sigma\ell m}\tilde{C}_{\ell m}\left[\gamma_{MM,\ell}^{N+1}\mathbf{M}_{\sigma\ell m}^{(3)}(k_{N+1})\mathbf{M}_{\sigma\ell m}'^{(3)}(k_{N+1}) + \chi_{NN,\ell}^{N+1}\mathbf{N}_{\sigma\ell m}^{(3)}(k_{N+1})\mathbf{N}_{\sigma\ell m}'^{(3)}(k_{N+1})\right]$$

$$+\frac{ik_{N+1}}{c}\sum_{\sigma\ell m}\tilde{C}_{\ell m}\left[\gamma_{NM,\ell}^{N+1}\mathbf{N}_{\sigma\ell m}^{(3)}(k_{N+1})\mathbf{M}_{\sigma\ell m}'^{(3)}(k_{N+1}) + \chi_{MN,\ell}^{N+1}\mathbf{M}_{\sigma\ell m}^{(3)}(k_{N+1})\mathbf{N}_{\sigma\ell m}'^{(3)}(k_{N+1})\right]$$

, (9)

$$\mathbf{G}_{es}^{(i,N+1)}(\mathbf{r},\mathbf{r}') = \frac{ik_{N+1}}{c}\sum_{\sigma\ell m}\tilde{C}_{\ell m}\left[\alpha_{MM,\ell}^{i}\mathbf{M}_{\sigma\ell m}^{(1)}(k_i)\mathbf{M}_{\sigma\ell m}'^{(3)}(k_{N+1}) + \beta_{NN,\ell}^{i}\mathbf{N}_{\sigma\ell m}^{(1)}(k_i)\mathbf{N}_{\sigma\ell m}'^{(3)}(k_{N+1})\right.$$
$$\left.+\gamma_{MM,\ell}^{i}\mathbf{M}_{\sigma\ell m}^{(3)}(k_i)\mathbf{M}_{\sigma\ell m}'^{(3)}(k_{N+1}) + \chi_{NN,\ell}^{i}\mathbf{N}_{\sigma\ell m}^{(3)}(k_i)\mathbf{N}_{\sigma\ell m}'^{(3)}(k_{N+1})\right]$$
$$+\frac{ik_{N+1}}{c}\sum_{\sigma\ell m}\tilde{C}_{\ell m}\left[\alpha_{NM,\ell}^{i}\mathbf{N}_{\sigma\ell m}^{(1)}(k_i)\mathbf{M}_{\sigma\ell m}'^{(3)}(k_{N+1}) + \beta_{MN,\ell}^{i}\mathbf{M}_{\sigma\ell m}^{(1)}(k_i)\mathbf{N}_{\sigma\ell m}'^{(3)}(k_{N+1})\right.$$
$$\left.+\gamma_{NM,\ell}^{i}\mathbf{N}_{\sigma\ell m}^{(3)}(k_i)\mathbf{M}_{\sigma\ell m}'^{(3)}(k_{N+1}) + \chi_{MN,\ell}^{i}\mathbf{M}_{\sigma\ell m}^{(3)}(k_i)\mathbf{N}_{\sigma\ell m}'^{(3)}(k_{N+1})\right], i = 2,3,...,N$$

(10)

$$\mathbf{G}_{es}^{(1,N+1)}(\mathbf{r},\mathbf{r}') = \frac{ik_{N+1}}{c}\sum_{\sigma\ell m}\tilde{C}_{\ell m}\left[\alpha_{MM,\ell}^{1}\mathbf{M}_{\sigma\ell m}^{(1)}(k_1)\mathbf{M}_{\sigma\ell m}'^{(3)}(k_{N+1}) + \beta_{NN,\ell}^{1}\mathbf{N}_{\sigma\ell m}^{(1)}(k_1)\mathbf{N}_{\sigma\ell m}'^{(3)}(k_{N+1})\right]$$
$$+\frac{ik_{N+1}}{c}\sum_{\sigma\ell m}\tilde{C}_{\ell m}\left[\alpha_{NM,\ell}^{1}\mathbf{N}_{\sigma\ell m}^{(1)}(k_1)\mathbf{M}_{\sigma\ell m}'^{(3)}(k_{N+1}) + \beta_{MN,\ell}^{1}\mathbf{M}_{\sigma\ell m}^{(1)}(k_1)\mathbf{N}_{\sigma\ell m}'^{(3)}(k_{N+1})\right]$$

, (11)

where $\{\alpha_{MM,\ell}^{i}, \beta_{NN,\ell}^{i}, \gamma_{MM,\ell}^{i}, \chi_{NN,\ell}^{i}\}$ are the conventional expansion coefficients, and $\{\alpha_{NM,\ell}^{i}, \beta_{MN,\ell}^{i}, \gamma_{NM,\ell}^{i}, \chi_{MN,\ell}^{i}\}$ are the additional expansion coefficients describing the mixing response of both the TE (described by $\mathbf{M}$) and TM (described by $\mathbf{N}$) waves, due to the axion coupling effect emerging exclusively from the discontinuity of $\Theta$-value across the boundary $r = R_i, i = 1,...,N$ [33,45].

According to the relation $\mathbf{E} = \frac{i\omega}{c}\int \mathbf{G}_e(\mathbf{r},\mathbf{r}')\bullet \mathbf{J}(\mathbf{r}')d^3\mathbf{r}'$ where $\mathbf{J}$ is the current source, with the application of Eq. (3), the appropriate boundary conditions for $\mathbf{G}_e^{(f,N+1)}$ are:

$$\mathbf{e}_r \times \mathbf{G}_e^{(i+1,N+1)}(\mathbf{r},\mathbf{r}') = \mathbf{e}_r \times \mathbf{G}_e^{(i,N+1)}(\mathbf{r},\mathbf{r}')$$
$$\frac{c}{i\omega}\mathbf{e}_r \times \nabla \times \mathbf{G}_e^{(i+1,N+1)}(\mathbf{r},\mathbf{r}') - \frac{c}{i\omega}\mathbf{e}_r \times \nabla \times \mathbf{G}_e^{(i,N+1)}(\mathbf{r},\mathbf{r}') = \frac{\alpha}{\pi}(\Theta_{i+1} - \Theta_i)\mathbf{e}_r \times \mathbf{G}_e^{(i,N+1)}(\mathbf{r},\mathbf{r}')$$

(12)

where $|\mathbf{r}| = R_i, i = 1,2,...,N$. Substituting Eq. (7), Eqs. (9)-(11) into Eq. (12), we obtain the following results:

$$j_\ell(\rho_{i+1,i})\delta_{i,N} + \alpha_{MM,\ell}^{i+1} j_\ell(\rho_{i+1,i}) + \gamma_{MM,\ell}^{i+1} h_\ell^{(1)}(\rho_{i+1,i}) = \alpha_{MM,\ell}^{i} j_\ell(\rho_{i,i}) + \gamma_{MM,\ell}^{i} h_\ell^{(1)}(\rho_{i,i}), \quad (13)$$

$$\beta_{MN,\ell}^{i+1} j_\ell(\rho_{i+1,i}) + \chi_{MN,\ell}^{i+1} h_\ell^{(1)}(\rho_{i+1,i}) = \beta_{MN,\ell}^{i} j_\ell(\rho_{i,i}) + \chi_{MN,\ell}^{i} h_\ell^{(1)}(\rho_{i,i}), \quad (14)$$

$$\xi_\ell^{(1)}(\rho_{i+1,i})\delta_{i,N} + \beta_{NN,\ell}^{i+1} \xi_\ell^{(1)}(\rho_{i+1,i}) + \chi_{NN,\ell}^{i+1} \xi_\ell^{(3)}(\rho_{i+1,i}) = \beta_{NN,\ell}^{i} \xi_\ell^{(1)}(\rho_{i,i}) + \chi_{NN,\ell}^{i} \xi_\ell^{(3)}(\rho_{i,i}),$$

$$(15)$$

$$\alpha_{NM,\ell}^{i+1} \xi_\ell^{(1)}(\rho_{i+1,i}) + \gamma_{NM,\ell}^{i+1} \xi_\ell^{(3)}(\rho_{i+1,i}) = \alpha_{NM,\ell}^{i} \xi_\ell^{(1)}(\rho_{i,i}) + \gamma_{NM,\ell}^{i} \xi_\ell^{(3)}(\rho_{i,i}), \quad (16)$$

$$\frac{ck_{i+1}}{i\omega}\left[\xi_\ell^{(1)}(\rho_{i+1,i})\delta_{i,N} + \alpha_{MM,\ell}^{i+1}\xi_\ell^{(1)}(\rho_{i+1,i}) + \gamma_{MM,\ell}^{i+1}\xi_\ell^{(3)}(\rho_{i+1,i})\right] - \frac{ck_i}{i\omega}\left[\alpha_{MM,\ell}^{i}\xi_\ell^{(1)}(\rho_{i,i}) + \gamma_{MM,\ell}^{i}\xi_\ell^{(3)}(\rho_{i,i})\right]$$
$$= \frac{\alpha}{\pi}(\Theta_{i+1} - \Theta_i)\left[\alpha_{NM,\ell}^{i}\xi_\ell^{(1)}(\rho_{i,i}) + \gamma_{NM,\ell}^{i}\xi_\ell^{(3)}(\rho_{i,i})\right]$$

$$, \quad (17)$$

$$\frac{ck_{i+1}}{i\omega}\left[\beta_{MN,\ell}^{i+1}\xi_\ell^{(1)}(\rho_{i+1,i}) + \chi_{MN,\ell}^{i+1}\xi_\ell^{(3)}(\rho_{i+1,i})\right] - \frac{ck_i}{i\omega}\left[\beta_{MN,\ell}^{i}\xi_\ell^{(1)}(\rho_{i,i}) + \chi_{MN,\ell}^{i}\xi_\ell^{(3)}(\rho_{i,i})\right]$$
$$= \frac{\alpha}{\pi}(\Theta_{i+1} - \Theta_i)\left[\beta_{NN,\ell}^{i}\xi_\ell^{(1)}(\rho_{i,i}) + \chi_{NN,\ell}^{i}\xi_\ell^{(3)}(\rho_{i,i})\right]$$

$$, \quad (18)$$

$$\frac{ck_{i+1}}{i\omega}\left[j_\ell(\rho_{i+1,i})\delta_{i,N} + \beta_{NN,\ell}^{i+1} j_\ell(\rho_{i+1,i}) + \chi_{NN,\ell}^{i+1} h_\ell^{(1)}(\rho_{i+1,i})\right] - \frac{ck_i}{i\omega}\left[\beta_{NN,\ell}^{i} j_\ell(\rho_{i,i}) + \chi_{NN,\ell}^{i} h_\ell^{(1)}(\rho_{i,i})\right]$$
$$= \frac{\alpha}{\pi}(\Theta_{i+1} - \Theta_i)\left[\beta_{MN,\ell}^{i} j_\ell(\rho_{i,i}) + \chi_{MN,\ell}^{i} h_\ell^{(1)}(\rho_{i,i})\right]$$

$$, \quad (19)$$

$$\frac{ck_{i+1}}{i\omega}\left[\alpha_{NM,\ell}^{i+1} j_\ell(\rho_{i+1,i}) + \gamma_{NM,\ell}^{i+1} h_\ell^{(1)}(\rho_{i+1,i})\right] - \frac{ck_i}{i\omega}\left[\alpha_{NM,\ell}^{i} j_\ell(\rho_{i,i}) + \gamma_{NM,\ell}^{i} h_\ell^{(1)}(\rho_{i,i})\right]$$
$$= \frac{\alpha}{\pi}(\Theta_{i+1} - \Theta_i)\left[\alpha_{MM,\ell}^{i} j_\ell(\rho_{i,i}) + \gamma_{MM,\ell}^{i} h_\ell^{(1)}(\rho_{i,i})\right]$$

$$, \quad (20)$$

with $\alpha_{MM,\ell}^{N+1} = \beta_{NN,\ell}^{N+1} = 0$, $\gamma_{MM,\ell}^{1} = \chi_{NN,\ell}^{1} = 0$, $\alpha_{NM,\ell}^{N+1} = \beta_{MN,\ell}^{N+1} = 0$, $\gamma_{NM,\ell}^{1} = \chi_{MN,\ell}^{1} = 0$

and $\delta_{i,N} = \begin{cases} 1, i = N \\ 0, i \neq N \end{cases}$. Moreover, $\xi_\ell^{(1)}(\rho_{i+1,i}) = \frac{d[\rho_{i+1,i} j_\ell(\rho_{i+1,i})]}{\rho_{i+1,i} d\rho_{i+1,i}}$,

$\xi_\ell^{(3)}(\rho_{i+1,i}) = \frac{d[\rho_{i+1,i} h_\ell^{(1)}(\rho_{i+1,i})]}{\rho_{i+1,i} d\rho_{i+1,i}}$, $\xi_\ell^{(1)}(\rho_{i,i}) = \frac{d[\rho_{i,i} j_\ell(\rho_{i,i})]}{\rho_{i,i} d\rho_{i,i}}$ and

$\xi_\ell^{(3)}(\rho_{i,i}) = \dfrac{d\left[\rho_{i,i} h_\ell^{(1)}(\rho_{i,i})\right]}{\rho_{i,i} d\rho_{i,i}}$ where two different types of size parameters $\rho_{i+1,i} = k_{i+1} R_i$ and $\rho_{i,i} = k_i R_i$ are defined.

To check the analytical results in the above derivation, we consider a simple case $\Theta_1 = \Theta_2 = ... = \Theta_N = 0$. For this case, the additional expansion coefficients reduce to zero and the scattering electric DGFs (Eqs. (9)-(11)) become the same as those in the literature [46-47]. In addition, we use the above results to obtain the electric DGFs of a TI sphere analytically as shown in Appendix C.

*(iii) Solving unknown expansion coefficients*

Applying the transfer matrix method similar to Moroz [50], we rewrite Eqs. (13)-(20) in the following matrix form:

$$\mathbf{\Psi}_\ell^{i+1} = \mathbf{T}_\ell^i \mathbf{\Psi}_\ell^i + \left(\mathbf{A}_\ell^i\right)^{-1} \mathbf{C}_\ell^i, \tag{21}$$

where $\mathbf{\Psi}_\ell^i, i = 1, 2, ..., N+1$ is defined as follows:

$$\mathbf{\Psi}_\ell^i = \left(\alpha_{MM,\ell}^i \quad \beta_{NN,\ell}^i \quad \gamma_{MM,\ell}^i \quad \chi_{NN,\ell}^i \quad \alpha_{NM,\ell}^i \quad \beta_{MN,\ell}^i \quad \gamma_{NM,\ell}^i \quad \chi_{MN,\ell}^i\right)^t, \tag{22}$$

where the superscript $t$ is a transport. $\mathbf{T}_\ell^i$ is the forward transfer matrix. The details of three matrices $\mathbf{T}_\ell^i$, $\mathbf{A}_\ell^i$ and $\mathbf{C}_\ell^i$ are expressed in Appendix D. Since the source is put into the outermost region as illustrated in Fig. 2, we have $\mathbf{C}_\ell^{N-1} = \mathbf{C}_\ell^{N-2} = ... = \mathbf{C}_\ell^1 = 0$. Hence, the correlation between the outermost $\mathbf{\Psi}_\ell^{N+1}$ and the innermost $\mathbf{\Psi}_\ell^1$ is

$$\begin{aligned}
\mathbf{\Psi}_\ell^{N+1} &= \mathbf{T}_\ell^N \mathbf{\Psi}_\ell^N + \left(\mathbf{A}_\ell^N\right)^{-1} \mathbf{C}_\ell^N = \mathbf{T}_\ell^N \left[\mathbf{T}_\ell^{N-1} \mathbf{\Psi}_\ell^{N-1} + \left(\mathbf{A}_\ell^{N-1}\right)^{-1} \mathbf{C}_\ell^{N-1}\right] + \left(\mathbf{A}_\ell^N\right)^{-1} \mathbf{C}_\ell^N \\
&= \mathbf{T}_\ell^N \mathbf{T}_\ell^{N-1} \mathbf{\Psi}_\ell^{N-1} + \left(\mathbf{A}_\ell^N\right)^{-1} \mathbf{C}_\ell^N = \mathbf{T}_\ell^N \mathbf{T}_\ell^{N-1} \left[\mathbf{T}_\ell^{N-2} \mathbf{\Psi}_\ell^{N-2} + \left(\mathbf{A}_\ell^{N-2}\right)^{-1} \mathbf{C}_\ell^{N-2}\right] + \left(\mathbf{A}_\ell^N\right)^{-1} \mathbf{C}_\ell^N \\
&= \mathbf{T}_\ell^N \mathbf{T}_\ell^{N-1} \mathbf{T}_\ell^{N-2} \mathbf{\Psi}_\ell^{N-2} + \left(\mathbf{A}_\ell^N\right)^{-1} \mathbf{C}_\ell^N = ... \\
&\vdots \\
&= \mathbf{T}_\ell^N \mathbf{T}_\ell^{N-1} \mathbf{T}_\ell^{N-2} \cdots \mathbf{T}_\ell^1 \mathbf{\Psi}_\ell^1 + \left(\mathbf{A}_\ell^N\right)^{-1} \mathbf{C}_\ell^N \\
&\equiv \mathbf{T}_\ell \mathbf{\Psi}_\ell^1 + \mathbf{D}_\ell
\end{aligned}$$

(23)

where $\mathbf{T}_\ell \equiv \mathbf{T}_\ell^N \mathbf{T}_\ell^{N-1} \cdots \mathbf{T}_\ell^1$ and $\mathbf{D}_\ell = \left(\mathbf{A}_\ell^N\right)^{-1} \mathbf{C}_\ell^N$. With $\alpha_{MM,\ell}^{N+1} = \beta_{NN,\ell}^{N+1} = 0$, $\gamma_{MM,\ell}^1 = \chi_{NN,\ell}^1 = 0$, $\alpha_{NM,\ell}^{N+1} = \beta_{MN,\ell}^{N+1} = 0$, $\gamma_{NM,\ell}^1 = \chi_{MN,\ell}^1 = 0$, and both the elements $T_{\alpha\beta}$ and $D_\alpha$ of matrices $\mathbf{T}_\ell$ and $\mathbf{D}_\ell$ respectively, Eq. (21) becomes:

$$\begin{cases} T_{11}\alpha_{MM,\ell}^1 + T_{12}\beta_{NN,\ell}^1 + T_{15}\alpha_{NM,\ell}^1 + T_{16}\beta_{MN,\ell}^1 = -D_1 \\ T_{21}\alpha_{MM,\ell}^1 + T_{22}\beta_{NN,\ell}^1 + T_{25}\alpha_{NM,\ell}^1 + T_{26}\beta_{MN,\ell}^1 = -D_2 \\ T_{51}\alpha_{MM,\ell}^1 + T_{52}\beta_{NN,\ell}^1 + T_{55}\alpha_{NM,\ell}^1 + T_{56}\beta_{MN,\ell}^1 = -D_5 \\ T_{61}\alpha_{MM,\ell}^1 + T_{62}\beta_{NN,\ell}^1 + T_{65}\alpha_{NM,\ell}^1 + T_{66}\beta_{MN,\ell}^1 = -D_6 \end{cases} \quad (24)$$

and it is straightforward to obtain four coefficients $\alpha_{MM,\ell}^1$, $\beta_{NN,\ell}^1$, $\alpha_{NM,\ell}^1$, $\beta_{MN,\ell}^1$. In addition, it is clear that $\mathbf{\Psi}_\ell^{i+1}$ can be determined when $\mathbf{\Psi}_\ell^i$ is known for $i = 1, 2, \ldots, N$ from Eq. (21). Finally, we obtain four coefficients $\gamma_{MM,\ell}^{N+1}$, $\chi_{NN,\ell}^{N+1}$, $\gamma_{NM,\ell}^{N+1}$, $\chi_{MN,\ell}^{N+1}$ in the following form:

$$\begin{cases} \gamma_{MM,\ell}^{N+1} = T_{31}\alpha_{MM,\ell}^1 + T_{32}\beta_{NN,\ell}^1 + T_{35}\alpha_{NM,\ell}^1 + T_{36}\beta_{MN,\ell}^1 + D_3 \\ \chi_{NN,\ell}^{N+1} = T_{41}\alpha_{MM,\ell}^1 + T_{42}\beta_{NN,\ell}^1 + T_{45}\alpha_{NM,\ell}^1 + T_{46}\beta_{MN,\ell}^1 + D_4 \\ \gamma_{NM,\ell}^{N+1} = T_{71}\alpha_{MM,\ell}^1 + T_{72}\beta_{NN,\ell}^1 + T_{75}\alpha_{NM,\ell}^1 + T_{76}\beta_{MN,\ell}^1 + D_7 \\ \chi_{MN,\ell}^{N+1} = T_{81}\alpha_{MM,\ell}^1 + T_{82}\beta_{NN,\ell}^1 + T_{85}\alpha_{NM,\ell}^1 + T_{86}\beta_{MN,\ell}^1 + D_8 \end{cases} \quad (25)$$

Hence all unknown coefficients $\mathbf{\Psi}_\ell^i$ with all $i = 1, 2, \ldots, N$ are solved.

By the way, we can use the alternative method to solve $\mathbf{\Psi}_\ell^i, i = 1, 2, \ldots, N$. After collecting Eqs. (13)-(20) with all $i = 1, 2, \ldots, N$ into the matrix equation $\mathbf{P}_\ell \mathbf{\Psi}_\ell = \mathbf{Q}_\ell$, where the size of the known matrix $\mathbf{P}_\ell$ is $8N \times 8N$, $\mathbf{Q}_\ell$ is $8N \times 1$, and the unknown matrix $\mathbf{\Psi}_\ell$ is $8N \times 1$ respectively, it is straightforward to obtain all unknown coefficients via $\mathbf{\Psi}_\ell = \mathbf{P}_\ell^{-1} \mathbf{Q}_\ell$.

**An application: a dipole in the vicinity of a TI stratified sphere**

Here we use the DGFs to formulate light scattering from a dipole in the vicinity of a TI stratified sphere. This dipole is placed at the position $z = d$ with its dipole moment $p$. We consider both the vertical and the horizontal orientations of this dipole as depicted in Fig. 2.

*(i) Vertical dipole*

We start with a point current source due to an electric dipole as the form $\mathbf{J}(\mathbf{r}) = -i\omega p \delta(\mathbf{r} - d\mathbf{e}_z)\mathbf{e}_z$, where $p$ is the dipole moment and $\mathbf{e}_z$ is the unit vector along the z-axis. According to $\mathbf{E} = \frac{i\omega}{c}\int \mathbf{G}_e(\mathbf{r},\mathbf{r}') \cdot \mathbf{J}(\mathbf{r}') d^3\mathbf{r}'$, the electric fields are:

$$\mathbf{E}(\mathbf{r}) = \frac{\omega^3 i p \sqrt{\varepsilon_{N+1}}}{c^3 k_{N+1} d} \sum_{\ell=1}^{\infty} (2\ell+1) \left[ j_\ell(k_{N+1}d) \mathbf{N}_{e\ell 0}^{(3)}(k_{N+1}) + \chi_{NN,\ell}^{N+1} h_\ell^{(1)}(k_{N+1}d) \mathbf{N}_{e\ell 0}^{(3)}(k_{N+1}) \right.$$
$$\left. + \chi_{MN,\ell}^{N+1} h_\ell^{(1)}(k_{N+1}d) \mathbf{M}_{e\ell 0}^{(3)}(k_{N+1}) \right] \tag{26}$$

for $d < r$,

$$\mathbf{E}(\mathbf{r}) = \frac{\omega^3 i p \sqrt{\varepsilon_{N+1}}}{c^3 k_{N+1} d} \sum_{\ell=1}^{\infty} (2\ell+1) h_\ell^{(1)}(k_{N+1}d) \left[ \mathbf{N}_{e\ell 0}^{(1)}(k_{N+1}) + \chi_{NN,\ell}^{N+1} \mathbf{N}_{e\ell 0}^{(3)}(k_{N+1}) + \chi_{MN,\ell}^{N+1} \mathbf{M}_{e\ell 0}^{(3)}(k_{N+1}) \right] \tag{27}$$

for $R_N < r < d$,

$$\mathbf{E}(\mathbf{r}) = \frac{\omega^3 i p \sqrt{\varepsilon_{N+1}}}{c^3 k_{N+1} d} \sum_{\ell=1}^{\infty} (2\ell+1) h_\ell^{(1)}(k_{N+1}d) \left[ \beta_{NN,\ell}^i \mathbf{N}_{e\ell 0}^{(1)}(k_i) + \chi_{NN,\ell}^i \mathbf{N}_{e\ell 0}^{(3)}(k_i) + \beta_{MN,\ell}^i \mathbf{M}_{e\ell 0}^{(1)}(k_i) \right.$$
$$\left. + \chi_{MN,\ell}^i \mathbf{M}_{e\ell 0}^{(3)}(k_i) \right] \tag{28}$$

for $R_{i-1} < r < R_i, i = 2, 3, ..., N$, and

$$\mathbf{E}(\mathbf{r}) = \frac{\omega^3 i p \sqrt{\varepsilon_{N+1}}}{c^3 k_{N+1} d} \sum_{\ell=1}^{\infty} (2\ell+1) h_\ell^{(1)}(k_{N+1}d) \left[ \beta_{NN,\ell}^1 \mathbf{N}_{e\ell 0}^{(1)}(k_1) + \beta_{MN,\ell}^1 \mathbf{M}_{e\ell 0}^{(1)}(k_1) \right] \tag{29}$$

for $r < R_1$, respectively.

Next we consider the reflected electric field excluding itself contribution at the dipole position in the following form:

$$E_\perp(d\mathbf{e}_z) = \frac{\omega^3 i p \sqrt{\varepsilon_{N+1}}}{c^3 k_{N+1}^2 d^2} \sum_{\ell=1}^{\infty} \ell(\ell+1)(2\ell+1) \chi_{NN,\ell}^{N+1} \left[ h_\ell^{(1)}(k_{N+1}d) \right]^2, \tag{30}$$

where the subscript $\perp$ denotes the z-component of the electric field. Assume that the medium in the outermost region is non-absorbing ($\text{Im}\,\varepsilon_{N+1} = 0$). Therefore the normalized decay rate of an emitting dipole near a TI stratified sphere is [51-52]:

$$\hat{b} = 1 + \frac{3\varepsilon_{N+1}}{2k_{N+1}^3} \text{Im}\left[\frac{E_\perp(d\mathbf{e}_z)}{p}\right]$$

$$= 1 + \frac{3}{2}\text{Re}\left(\sum_{\ell=1}^{\infty} \ell(\ell+1)(2\ell+1) \chi_{NN,\ell}^{N+1} \left[\frac{h_\ell^{(1)}(k_{N+1}d)}{k_{N+1}d}\right]^2\right), \quad (31)$$

which is only related to the conventional expansion coefficient $\chi_{NN,\ell}^{N+1}$ but not to the additional expansion coefficients [34-35]. Hence the axion coupling effect affects the results obtained from Eq. (31) via the coefficient $\chi_{NN,\ell}^{N+1}$ as solved within the framework of the DGF formulation.

*(ii) Horizontal dipole*

In a similar way as in (i), with the current density $\mathbf{J}(\mathbf{r}) = -i\omega p \delta(\mathbf{r} - d\mathbf{e}_z)\mathbf{e}_x$ where $\mathbf{e}_x$ is the unit vector along the x-axis, The electric fields are:

$$\mathbf{E}(\mathbf{r}) = -\frac{\omega^3 i p \sqrt{\varepsilon_{N+1}}}{c^3} \sum_{\ell=1}^{\infty} \frac{(2\ell+1)}{\ell(\ell+1)} \left\{\left[j_\ell(k_{N+1}d) + h_\ell^{(1)}(k_{N+1}d)\gamma_{MM,\ell}^{N+1}\right]\mathbf{M}_{o\ell 1}^{(3)}(k_{N+1}) + h_\ell^{(1)}(k_{N+1}d)\gamma_{NM,\ell}^{N+1}\mathbf{N}_{o\ell 1}^{(3)}(k_{N+1})\right\}$$

$$-\frac{\omega^3 i p \sqrt{\varepsilon_{N+1}}}{c^3} \sum_{\ell=1}^{\infty} \frac{(2\ell+1)}{\ell(\ell+1)} \left\{\left[\xi_\ell^{(1)}(k_{N+1}d) + \xi_\ell^{(3)}(k_{N+1}d)\chi_{NN,\ell}^{N+1}\right]\mathbf{N}_{e\ell 1}^{(3)}(k_{N+1}) + \xi_\ell^{(3)}(k_{N+1}d)\chi_{MN,\ell}^{N+1}\mathbf{M}_{e\ell 1}^{(3)}(k_{N+1})\right\}$$

(32)

for $d < r$,

$$\mathbf{E}(\mathbf{r}) = -\frac{\omega^3 i p \sqrt{\varepsilon_{N+1}}}{c^3} \sum_{\ell=1}^{\infty} \frac{(2\ell+1)}{\ell(\ell+1)} \left\{h_\ell^{(1)}(k_{N+1}d)\left[\mathbf{M}_{o\ell 1}^{(1)}(k_{N+1}) + \gamma_{MM,\ell}^{N+1}\mathbf{M}_{o\ell 1}^{(3)}(k_{N+1})\right] + h_\ell^{(1)}(k_{N+1}d)\gamma_{NM,\ell}^{N+1}\mathbf{N}_{o\ell 1}^{(3)}(k_{N+1})\right\}$$

$$-\frac{\omega^3 i p \sqrt{\varepsilon_{N+1}}}{c^3} \sum_{\ell=1}^{\infty} \frac{(2\ell+1)}{\ell(\ell+1)} \left\{\xi_\ell^{(3)}(k_{N+1}d)\left[\mathbf{N}_{e\ell 1}^{(1)}(k_{N+1}) + \chi_{NN,\ell}^{N+1}\mathbf{N}_{e\ell 1}^{(3)}(k_{N+1})\right] + \xi_\ell^{(3)}(k_{N+1}d)\chi_{MN,\ell}^{N+1}\mathbf{M}_{e\ell 1}^{(3)}(k_{N+1})\right\}$$

(33)

for $R_N < r < d$,

$$\mathbf{E}(\mathbf{r}) = -\frac{\omega^3 i p \sqrt{\varepsilon_{N+1}}}{c^3} \sum_{\ell=1}^{\infty} \frac{(2\ell+1)}{\ell(\ell+1)} \left[\alpha_{MM,\ell}^i h_\ell^{(1)}(k_{N+1}d)\mathbf{M}_{o\ell 1}^{(1)}(k_i) + \beta_{NN,\ell}^i \xi_\ell^{(3)}(k_{N+1}d)\mathbf{N}_{e\ell 1}^{(1)}(k_i)\right.$$

$$\left.+\gamma_{MM,\ell}^i h_\ell^{(1)}(k_{N+1}d)\mathbf{M}_{o\ell 1}^{(3)}(k_i) + \chi_{NN,\ell}^i \xi_\ell^{(3)}(k_{N+1}d)\mathbf{N}_{e\ell 1}^{(3)}(k_i)\right]$$

$$-\frac{\omega^3 i p \sqrt{\varepsilon_{N+1}}}{c^3} \sum_{\ell=1}^{\infty} \frac{(2\ell+1)}{\ell(\ell+1)} \left[\alpha_{NM,\ell}^i h_\ell^{(1)}(k_{N+1}d)\mathbf{N}_{o\ell 1}^{(1)}(k_i) + \beta_{MN,\ell}^i \xi_\ell^{(3)}(k_{N+1}d)\mathbf{M}_{e\ell 1}^{(1)}(k_i)\right.$$

$$\left.+\gamma_{NM,\ell}^i h_\ell^{(1)}(k_{N+1}d)\mathbf{N}_{o\ell 1}^{(3)}(k_i) + \chi_{MN,\ell}^i \xi_\ell^{(3)}(k_{N+1}d)\mathbf{M}_{e\ell 1}^{(3)}(k_i)\right]$$

(34)

for $R_{i-1} < r < R_i, i = 2, 3, ..., N$, and

$$\mathbf{E}(\mathbf{r}) = -\frac{\omega^3 ip\sqrt{\varepsilon_{N+1}}}{c^3} \sum_{\ell=1}^{\infty} \frac{(2\ell+1)}{\ell(\ell+1)} \left[ \alpha_{MM,\ell}^1 h_\ell^{(1)}(k_{N+1}d) \mathbf{M}_{o\ell 1}^{(1)}(k_1) + \beta_{NN,\ell}^1 \xi_\ell^{(3)}(k_{N+1}d) \mathbf{N}_{e\ell 1}^{(1)}(k_1) \right]$$

$$-\frac{\omega^3 ip\sqrt{\varepsilon_{N+1}}}{c^3} \sum_{\ell=1}^{\infty} \frac{(2\ell+1)}{\ell(\ell+1)} \left[ \alpha_{NM,\ell}^1 h_\ell^{(1)}(k_{N+1}d) \mathbf{N}_{o\ell 1}^{(1)}(k_1) + \beta_{MN,\ell}^1 \xi_\ell^{(3)}(k_{N+1}d) \mathbf{M}_{e\ell 1}^{(1)}(k_1) \right]$$

(35)

for $r < R_1$, respectively.

Furthermore, the reflected electric field at the dipole position is:

$$E_\parallel(d\mathbf{e}_z) = \frac{\omega^3 ip\sqrt{\varepsilon_{N+1}}}{2c^3} \sum_{\ell=1}^{\infty} (2\ell+1) \left\{ \gamma_{MM,\ell}^{N+1} \left[ h_\ell^{(1)}(k_{N+1}d) \right]^2 + \chi_{NN,\ell}^{N+1} \left[ \xi_\ell^{(3)}(k_{N+1}d) \right]^2 \right\}, \quad (36)$$

where the subscript ∥ denotes the x-component of the electric field. For the non-absorbing medium in the outermost region, we obtain the normalized decay rate in the horizontal case as follows [51-52]:

$$\hat{b} = 1 + \frac{3}{4} \text{Re} \left( \sum_{\ell=1}^{\infty} (2\ell+1) \left\{ \gamma_{MM,\ell}^{N+1} \left[ h_\ell^{(1)}(k_{N+1}d) \right]^2 + \chi_{NN,\ell}^{N+1} \left[ \xi_\ell^{(3)}(k_{N+1}d) \right]^2 \right\} \right), \quad (37)$$

which is only related to the conventional expansion coefficients $\gamma_{MM,\ell}^{N+1}$ and $\chi_{NN,\ell}^{N+1}$ but not to the additional expansion coefficients [34-35]. Hence the axion coupling effect affects the results obtained from Eq. (37) via both $\gamma_{MM,\ell}^{N+1}$ and $\chi_{NN,\ell}^{N+1}$, as solved with the help of the DGF formulation.

**Numerical results and discussions**

In this section, we apply our above results to some numerical calculation of the TME-modified decay rate spectrum for an emitting dipole in the vicinity of a stratified TI sphere. For a metal (Ag) shell, the dielectric function is described by the Drude model [21,27]:

$$\varepsilon_{Ag} = 1 - \frac{\omega_p^2}{\omega^2 + i\omega\Gamma}, \quad (38)$$

with the plasmon frequency $\omega_p = 1.36 \times 10^{16}$ rad/s and the damping constant $\Gamma = \Gamma_{Bulk} + \frac{\alpha v_F}{t_{Ag}}$ where $t_{Ag}$ is the thickness of an Ag material, $\Gamma_{Bulk} = 2.56 \times 10^{13}$ Hz, $v_F = 1.39 \times 10^6$ m/s and choosing $\alpha = 1$. Throughout this paper, the dipole moment is taken to be $p = 1$. Moreover, the axion parameter is chosen in the form $\Theta = (2n+1)\pi$, and $n$ is an integer describing TI on the TME effect with the broken time-reversal symmetry. In the following, we first consider TI materials to be non-absorbed. Next, the losses of TI materials are taken into account.

*(i) TI materials without the effect of losses*

First we choose the dielectric function of a TI core to be that of TiBiSe$_2$ in the following form [21,27,53]:

$$\varepsilon_{TI} = 1 - \frac{\omega_e^2}{\omega^2 - \omega_R^2 + i\omega\Gamma_R} \approx 1 - \frac{\omega_e^2}{\omega^2 - \omega_R^2}, \tag{39}$$

with $\omega_R = 1.68 \times 10^{12}$ rad/s and $\omega_e = \sqrt{3}\omega_R$. In this case, the damping constant $\Gamma_R$ is much smaller than the resonance frequency $\omega_R$ and we neglect that. Fig. 3 (a) shows a configuration of a TI sphere with a metal shell of inner radius $R_1$ and outer radius $R_2$. A dipole with two mutually orthogonal orientations (vertical and horizontal) is located at a radial distance $d$ on the z-axis outside the shell. We present the normalized decay rate as a function of the normalized emission frequency for a vertical (Fig. 3(b)) and a horizontal (Fig. 3(c)) dipole near a silver shell of small dimensions $R_1 = 8nm$ and $R_2 = 10nm$ in Fig. 3(a). The location of a dipole is at $d = 12nm$. To ensure our results are accurate, we compare them with the results obtained in the electrostatics approximation as presented in Ref. [21]. Note that the case with a non-TI core ($\Theta_1 = 0$) is set as a reference for the TME effect. In this case, it is found that the DGF formulation (in green) is equivalent to the electrostatics approximation (in black) for both vertical and horizontal dipoles, and so is another case with TI-core ($\Theta_1 = 101\pi$). Moreover, it is observed that the TME effect causes red-shifts significantly for the bonding modes of the metal shell with the lower order multipolar resonances as shown in Ref. [21].

Moreover, we focus on the system of larger dimensions $R_1 = 100nm$ and $R_2 = 120nm$ in Fig. 3(a). In a similar way as discussed above, we study on the TME-modified decay rate spectrum as shown in Fig. 4. Note that we use the DGF formulation to obtain accurate results especially at the low-frequency regimes. To ensure our numerical results are correct, we take the cutoff of the order of the multipolar mode to be $\ell_{max} = 50$. The dipole is located at $d = 125nm$, and the case with a non-TI core ($\Theta_1 = 0$, in black) is set as a reference for the TME effect in Fig. 4(a) and (b). For this case, as in analyzing the peaks of the decay rate spectrum in our previous works [21], we find various peaks in these two figures and have identified eight bonding modes ($\ell=1, \omega \sim 0.15\omega_p$; $\ell=2, \omega \sim 0.3\omega_p$; $\ell=3, \omega \sim 0.38\omega_p$; $\ell=4$, $\omega \sim 0.46\omega_p$; $\ell=5, \omega \sim 0.51\omega_p$; $\ell=6, \omega \sim 0.55\omega_p$; $\ell=7, \omega \sim 0.58\omega_p$; $\ell=8, \omega \sim 0.6\omega_p$) corresponding to different multipolar resonances, as manifested in Fig. 4(c) only for the vertical dipole. It is observed that the TME effect ($\Theta_1 = 101\pi$, in red) causes red-shifts for these low order bonding modes. Analytically, we can use both the electromagnetic theory and plasmonic hybridization model to demonstrate that bonding and antibonding modes for each order of multipolar resonances are all red-shifts [54]. However, for the higher order multipoles, it is observed that both bonding and antibonding modes are hardly distinguishable [21], as manifested in Fig. 4(c) for the order $\ell = 40$. Hence, the TME-induced red-shifts after a summation over these modes cannot be discriminated from the total decay rate especially at the high frequency regimes of Fig. 4(a) and (b). Moreover, the TME effect for vertical (Fig. 4(a)) and horizontal (Fig. 4(b)) dipoles is not sensitive to the dipole orientation. Furthermore, we show how the TME effect influences both the bonding and antibonding modes of four different multipolar orders: (a) $\ell = 5$, (b) $\ell = 10$, (c) $\ell = 15$, and (d) $\ell = 20$ only for the vertical dipole in Fig. 5. It is found that the TME effect leads to increasing red-shifts for the bonding modes and lower order antibonding modes, as well as decreasing amplitudes for the bonding modes and increasing ones for the lower order antibonding modes. However, for the antibonding modes of the higher order $\ell = 20$, it is found that both the increasing red-shifts and amplitudes are insignificant especially as $\Theta_1$ increases form $101\pi$ to $201\pi$.

In order to go beyond two material regions as presented in Ref. [21], we give an example of four material regions $N = 4$ in Fig. 2. A composite sphere consisting of a TI core and three adjoining concentric metal-TI-metal shells with four radii $R_1 = 50nm$, $R_2 = 60nm$, $R_3 = 110nm$ and $R_4 = 120nm$ is shown in Fig. 6 (a). The location of the dipole is at $d = 125nm$. To ensure our numerical results are correct, the cutoff of the order of multipolar mode $\ell_{max} = 70$ is needed. The case with both a non-TI core ($\Theta_1 = 0$) and a non-TI shell ($\Theta_3 = 0$) is set as a reference for the TME effect as shown in the black line of Fig. 6(b) and (c) for a vertical and a horizontal dipole, respectively. For this case, two bonding and two antibonding modes are found for each multipolar order $\ell$ according to the coupling effects of two plasmonic shells as presented in Fig. 6(d) only for the vertical dipole; one of two higher-order ($\ell \geq 5$) bonding modes at the high frequency regimes is too weak to observe and two higher-order ($\ell \geq 5$) antibonding modes merge with each other. Furthermore, we have identified eight bonding modes ($\ell=1$, $\omega \sim 0.148\omega_p$; $\ell=2$, $\omega \sim 0.249\omega_p$; $\ell=3$, $\omega \sim 0.319\omega_p$; $\ell=4$, $\omega \sim 0.372\omega_p$; $\ell=5$, $\omega \sim 0.413\omega_p$; $\ell=6$, $\omega \sim 0.446\omega_p$; $\ell=7$, $\omega \sim 0.473\omega_p$; $\ell=8$, $\omega \sim 0.497\omega_p$) corresponding to different multipolar resonances as manifested in Fig. 6(d). It is observed that the TME effect originated from the TI-shell ($\Theta_3 = 101\pi$, in blue) causes more significant modifications of the decay rate spectrum than one from the TI-core ($\Theta_1 = 101\pi$, in red), which can be understood that the distance between the dipole and the TI-shell is smaller than one between the dipole and the TI-core. As a similar phenomenological characteristic for the previous system of a metal-coated TI sphere as shown in Fig. 4(a) and (b), we find the observable TME-induced red-shifts from the TI-shell for these identified bonding modes. However, as illustrated in Fig. 6(d) for the order $\ell = 60$, both bonding and antibonding modes are hardly distinguishable. Hence, after a summation over all multipolar modes, the TME-induced red-shifts cannot be clearly identified from the total decay rate at the high frequency regimes ($\omega > 0.65\omega_p$). To observe how the TME effect affects both the bonding and antibonding modes of different multipolar orders, we give four examples: (a) $\ell = 5$, (b) $\ell = 10$, (c) $\ell = 20$ and (d) $\ell = 40$ as shown in Fig. 7 only for the vertical dipole via increasing the axion parameter $\Theta_3$ from 0 to $201\pi$ ($\Theta_1 = 0$ in this case). For $\ell = 5$, two antibonding modes are hardly distinguishable when $\Theta_3 = 201\pi$. Furthermore, it is observed that the TME effect leads to increasing red-shifts for the bonding modes and lower order antibonding modes, as well as decreasing amplitudes for the bonding modes and increasing ones for the lower order antibonding modes. However, for the antibonding modes of the higher order $\ell = 40$, both the increasing red-shifts and amplitudes are insignificant especially as $\Theta_3$ increases form $101\pi$ to $201\pi$.

Moreover, we present a complicated structure with ten material regions $N = 10$ in Fig. 2. An alternating metal-TI stratified sphere with ten radii $R_k = 10k, k = 1, 2, ..., 10$ (unit: nm) is illustrated in Fig. 8(a). The location of the dipole is at $d = 105nm$. To consider the contribution of higher multipolar modes, the cutoff of the order of multipolar mode $\ell_{max} = 70$ is needed. The case with a non-TI core ($\Theta_1 = 0$) and non-TI shells ($\Theta_{2k+1} = 0, k = 1, 2, 3, 4$) is set as a reference for the TME effect as shown in the black line of Fig. 8(b) and (c) for a vertical and a horizontal dipole, respectively. For this case, we find five bonding and five antibonding modes for each multipolar order $\ell$ according to the coupling effect of five plasmonic shells and these modes merge with each other when $\ell \geq 10$, as presented in Fig. 8(d) only for the vertical dipole. Moreover, the observable fluctuation at the high frequency regimes ($\omega > 0.8\omega_p$) is caused by the contribution of some antibonding modes as shown in Fig. 8(d). Furthermore, we have identified eight bonding modes ($\ell = 1$, $\omega \sim 0.096\omega_p$; $\ell = 1$, $\omega \sim 0.143\omega_p$; $\ell = 2$, $\omega \sim 0.165\omega_p$; $\ell = 3$, $\omega \sim 0.227\omega_p$; $\ell = 2$, $\omega \sim 0.239\omega_p$; $\ell = 4$, $\omega \sim 0.283\omega_p$; $\ell = 3$, $\omega \sim 0.318\omega_p$; $\ell = 5$, $\omega \sim 0.333\omega_p$) corresponding to different multipolar resonances as manifested in Fig. 8(d). It is observed that the TME effect originated from the outermost TI-shell ($\Theta_9 = 101\pi$, in blue) causes more significant modifications of the decay rate spectrum than one from other TI-core ($\Theta_i = 101\pi$: $i = 1$ in magenta, $i = 3$ in violet, $i = 5$ in green or $i = 7$ in red) as explained in the previous example for four material regions. Moreover, two identified bonding modes ($\ell = 3$, $\omega \sim 0.318\omega_p$; $\ell = 5$, $\omega \sim 0.333\omega_p$) merge as $\Theta_9 = 101\pi$. However, as illustrated in Fig. 8(d) for the order $\ell = 60$, both bonding and antibonding modes are hardly distinguishable. Hence, after a summation over all multipolar modes, the TME-induced red-shifts cannot be clearly identified from the total decay rate at $\omega > 0.4\omega_p$. To explain how the TME effect affects both the bonding and antibonding modes of different multipolar orders, we give four examples: (a) $\ell = 3$, (b) $\ell = 10$, (c) $\ell = 20$ and (d) $\ell = 30$ as shown in Fig. 9 only for the vertical dipole via increasing the axion parameter $\Theta_9$ from 0 to $201\pi$ ($\Theta_{2k+1} = 0, k = 0,1,2,3$ in this case). For $\ell = 3$, the increasing TME-induced red-shifts of five bonding and five antibonding modes are observed as $\Theta_9$ increases form 0 to $201\pi$. It is observed that the TME effect leads to increasing red-shifts for the bonding modes and lower order antibonding modes, as well as decreasing amplitudes for the bonding modes and increasing ones for the lower order antibonding modes. However, for the antibonding modes of the higher order $\ell = 30$, both the increasing red-shifts and amplitudes are insignificant especially as $\Theta_9$ increases form $101\pi$ to $201\pi$.

*(ii) TI materials with the effect of losses*

Next we consider another type of TI materials of which the dielectric function includes the damping term. We choose the dielectric function of a TI core to be that of the bulk $Bi_2Se_3$ with spin-orbital coupling via the density functional theory [42] as presented in Fig. 10(a) and recalculate the molecular decay rate for the system of dimensions $R_1 = 100nm$ and $R_2 = 120nm$ with a vertical dipole located at $d = 125nm$ in Fig. 3(a). To ensure our numerical results are correct, we take the cutoff of the order of the multipolar mode to be $\ell_{max} = 65$. The case with a non-TI core ($\Theta_1 = 0$, in black) is set as a reference for the TME effect in Fig. 10(b). For this case, we identified five antibonding modes ($\ell = 1, \omega \sim 0.141\omega_p$; $\ell = 2, \omega \sim 0.312\omega_p$; $\ell = 3, \omega \sim 0.44\omega_p$; $\ell = 4, \omega \sim 0.529\omega_p$; $\ell = 5, \omega \sim 0.586\omega_p$) corresponding to different multipolar resonances as manifested in Fig. 10(c). Unlike the behavior of Fig. 4(a), the bonding modes are too small to observe at the lower frequency region ($\omega < 0.2\omega_p$). The TME-induced red-shifts for these antibonding modes are insignificant since the dominance of bulk scattering [55-56] over surface scattering [26]. Furthermore, we show how the TME effect influences both the bonding and antibonding modes of two different multipolar orders $\ell = 5$ and $\ell = 10$ in Fig. 10(d) and (e). It is observed that the TME effect causes insignificant red-shifts for the bonding modes. Moreover, when $\Theta_1$ increases form 0 to $401\pi$, the increasing amplitudes for the bonding modes as well as decreasing amplitudes for the antibonding modes are observed. For the antibonding modes of the higher order $\ell = 10$, the TME-induced red-shifts are insignificant.

In addition, if the radius of the TI core is smaller than five quintuple layers (5QL), the surface (quantum) effect cannot be ignored due to the larger surface-to-volume ratio of that core. Hence, we replace the dielectric function of the bulk $Bi_2Se_3$ with one of the 5QL $Bi_2Se_3$ slab [42] via the density functional theory as presented in Fig. 11(a). We calculate the normalized decay rate of a vertical dipole near a metal-coated TI sphere of dimensions $R_1 = 5nm$ and $R_2 = 7nm$ in Fig. 3(a). This dipole is located at $d = 8nm$. To ensure our numerical results are correct, we take the cutoff of the order of the multipolar mode to be $\ell_{max} = 45$. The case with a non-TI core ($\Theta_1 = 0$, in black) is set as a reference for the TME effect in Fig. 11(b).

For this case, we only identified one antibonding mode ($\ell=1$, $\omega \sim 0.552\omega_p$) corresponding to different multipolar resonances as manifested in Fig. 11(c). Furthermore, we show the TME-induced modification for two different multipolar orders $\ell=5$ and $\ell=10$ in Fig. 11(d) and (e). It is found that the TME effect causes the observable red-shift for the antibonding mode of the order $\ell=5$ but insignificant one of the order $\ell=10$. Moreover, when $\Theta_1$ increases from 0 to $201\pi$, the decreasing amplitudes for the antibonding mode are observed for the order $\ell=5$ as presented in Fig. 11(d).

**Conclusion**

In this paper, we have constructed the dyadic Green's functions for a TI stratified sphere. Within the framework of axion electrodynamics, the additional expansion coefficients for the electric DGFs emerge exclusively from the discontinuity of the axion parameter across the interface of two adjacent layers. Moreover, we use the transfer matrix method to solve all unknown expansion coefficients, as well as these DGFs to establish the formulation of light scattering form a dipole near a TI stratified sphere. In the numerical studies, due to the TME effect, we have first observed the modifications of the decay rate spectrum for an emitting dipole near a TI sphere with a metal shell. For such a shell with small dimensions, the results obtained in the DGF formulation are equivalent to those calculated in the electrostatics approximations as shown in Ref. [21]. For the multipolar plasmonic resonances of a TI sphere with the metal shell, a metal-TI-metal-coated TI sphere and an alternating metal-TI stratified sphere, we find significant red-shifts for the bonding and lower order antibonding modes but insignificant ones for the higher order antibonding modes. Next, for a metal-coated TI sphere, we take into account the effects of losses in the TI core of which the dielectric function is chosen to be the form of the bulk or the 5QL slab and then some modifications of the TME-induced decay rate spectrum are obviously suppressed. These phenomenological findings provide us with some useful guidance of protecting the TME effect via molecular fluorescence experiments.

**Appendix A** The complete expressions of two spherical vector wave functions $\mathbf{M}$ and $\mathbf{N}$

The spherical vector wave function $\mathbf{M}$ is [46-47]:

$$\mathbf{M}^{(1)}_{o\ell m} = \mathbf{e}_\theta \frac{m}{\sin\theta} j_\ell(\rho) P_\ell^m(\cos\theta)\cos m\varphi - \mathbf{e}_\varphi j_\ell(\rho)\frac{\partial P_\ell^m(\cos\theta)}{\partial\theta}\sin m\varphi, \qquad \text{(A-1)}$$

$$\mathbf{M}^{(1)}_{e\ell m} = -\mathbf{e}_\theta \frac{m}{\sin\theta} j_\ell(\rho) P_\ell^m(\cos\theta)\sin m\varphi - \mathbf{e}_\varphi j_\ell(\rho)\frac{\partial P_\ell^m(\cos\theta)}{\partial\theta}\cos m\varphi, \qquad \text{(A-2)}$$

$$\mathbf{M}_{o\ell m}^{(3)} = \mathbf{e}_\theta \frac{m}{\sin\theta} h_\ell^{(1)}(\rho) P_\ell^m(\cos\theta)\cos m\varphi - \mathbf{e}_\varphi h_\ell^{(1)}(\rho)\frac{\partial P_\ell^m(\cos\theta)}{\partial\theta}\sin m\varphi, \qquad \text{(A-3)}$$

$$\mathbf{M}_{e\ell m}^{(3)} = -\mathbf{e}_\theta \frac{m}{\sin\theta} h_\ell^{(1)}(\rho) P_\ell^m(\cos\theta)\sin m\varphi - \mathbf{e}_\varphi h_\ell^{(1)}(\rho)\frac{\partial P_\ell^m(\cos\theta)}{\partial\theta}\cos m\varphi. \qquad \text{(A-4)}$$

The spherical vector wave function **N** is [46-47]:

$$\begin{aligned}\mathbf{N}_{e\ell m}^{(1)} &= \mathbf{e}_r \ell(\ell+1)\frac{j_\ell(\rho)}{\rho} P_\ell^m(\cos\theta)\cos m\varphi + \mathbf{e}_\theta \xi_\ell^{(1)}(\rho)\frac{\partial P_\ell^m(\cos\theta)}{\partial\theta}\cos m\varphi \\ &\quad -\mathbf{e}_\varphi \frac{m}{\sin\theta}\xi_\ell^{(1)}(\rho) P_\ell^m(\cos\theta)\sin m\varphi\end{aligned} \qquad \text{(A-5)}$$

$$\begin{aligned}\mathbf{N}_{o\ell m}^{(1)} &= \mathbf{e}_r \ell(\ell+1)\frac{j_\ell(\rho)}{\rho} P_\ell^m(\cos\theta)\sin m\varphi + \mathbf{e}_\theta \xi_\ell^{(1)}(\rho)\frac{\partial P_\ell^m(\cos\theta)}{\partial\theta}\sin m\varphi \\ &\quad +\mathbf{e}_\varphi \frac{m}{\sin\theta}\xi_\ell^{(1)}(\rho) P_\ell^m(\cos\theta)\cos m\varphi\end{aligned} \qquad \text{(A-6)}$$

$$\begin{aligned}\mathbf{N}_{e\ell m}^{(3)} &= \mathbf{e}_r \ell(\ell+1)\frac{h_\ell^{(1)}(\rho)}{\rho} P_\ell^m(\cos\theta)\cos m\varphi + \mathbf{e}_\theta \xi_\ell^{(3)}(\rho)\frac{\partial P_\ell^m(\cos\theta)}{\partial\theta}\cos m\varphi \\ &\quad -\mathbf{e}_\varphi \frac{m}{\sin\theta}\xi_\ell^{(3)}(\rho) P_\ell^m(\cos\theta)\sin m\varphi\end{aligned} \qquad \text{(A-7)}$$

$$\begin{aligned}\mathbf{N}_{o\ell m}^{(3)} &= \mathbf{e}_r \ell(\ell+1)\frac{h_\ell^{(1)}(\rho)}{\rho} P_\ell^m(\cos\theta)\sin m\varphi + \mathbf{e}_\theta \xi_\ell^{(3)}(\rho)\frac{\partial P_\ell^m(\cos\theta)}{\partial\theta}\sin m\varphi \\ &\quad +\mathbf{e}_\varphi \frac{m}{\sin\theta}\xi_\ell^{(3)}(\rho) P_\ell^m(\cos\theta)\cos m\varphi\end{aligned} \qquad \text{(A-8)}$$

where $\xi_\ell^{(1)}(\rho) = \dfrac{d[\rho j_\ell(\rho)]}{\rho d\rho}$ and $\xi_\ell^{(3)}(\rho) = \dfrac{d[\rho h_\ell^{(1)}(\rho)]}{\rho d\rho}$. The variable $\rho = kr$ is dimensionless and $\mathbf{e}_r, \mathbf{e}_\theta, \mathbf{e}_\varphi$ are three orthonormal vectors in spherical coordinates.

**Appendix B** Derivation of Eq. (7) via the method of $\mathbf{G}_m$

While the derivation of the conventional free-space electric DGF $\mathbf{G}_{e0}$ in spherical coordinates has been widely studied in the literature, it is still worthwhile to re-derive it because of ensuring its correctness in Gaussian units. We follow the method of $\mathbf{G}_m$ as shown in Tai [46] to first obtain the free-space magnetic DGF $\mathbf{G}_{m0}$ and later $\mathbf{G}_{e0}$ via the Ohm-Rayleigh method.

Considering a homogeneous dielectric response and axion field ($\nabla\Theta = 0$) in the free space, we have the conventional vector wave equation [46-47]:

$$\nabla\times\nabla\times\mathbf{G}_{e0}(\mathbf{r},\mathbf{r}') - k^2\mathbf{G}_{e0}(\mathbf{r},\mathbf{r}') = \frac{4\pi}{c}\mathbf{I}\delta(\mathbf{r}-\mathbf{r}'), \qquad \text{(B-1)}$$

with the wavenumber $k = \dfrac{\omega}{c}\sqrt{\varepsilon}$. $\mathbf{I}$ and $\delta(\mathbf{r}-\mathbf{r}')$ denote the unit dyadic and the Dirac delta function, respectively. Note that the relation between $\mathbf{G}_{m0}$ and $\mathbf{G}_{e0}$ is $\mathbf{G}_{m0}(\mathbf{r},\mathbf{r}') = \nabla \times \mathbf{G}_{e0}(\mathbf{r},\mathbf{r}')$ and hence Eq. (B-1) becomes:

$$\nabla \times \mathbf{G}_{m0}(\mathbf{r},\mathbf{r}') = k^2 \mathbf{G}_{e0}(\mathbf{r},\mathbf{r}') + \frac{4\pi}{c} \mathbf{I}\delta(\mathbf{r}-\mathbf{r}'). \tag{B-2}$$

Taking the curl of Eq. (B-2), we have the following wave equation of $\mathbf{G}_{m0}$:

$$\nabla \times \nabla \times \mathbf{G}_{m0}(\mathbf{r},\mathbf{r}') - k^2 \mathbf{G}_{m0}(\mathbf{r},\mathbf{r}') = \frac{4\pi}{c} \nabla \times \mathbf{I}\delta(\mathbf{r}-\mathbf{r}'). \tag{B-3}$$

According to the Ohm-Rayleigh method, we expand $\nabla \times \mathbf{I}\delta(\mathbf{r}-\mathbf{r}')$ as follows:

$$\nabla \times \mathbf{I}\delta(\mathbf{r}-\mathbf{r}') \equiv \int_0^\infty d\kappa \sum_{\sigma\ell m} \left[ \mathbf{N}_{\sigma\ell m}^{(1)}(\kappa) \mathbf{A}_{\sigma\ell m}(\kappa) + \mathbf{M}_{\sigma\ell m}^{(1)}(\kappa) \mathbf{B}_{\sigma\ell m}(\kappa) \right]. \tag{B-4}$$

with

$$\begin{aligned}\mathbf{A}_{\sigma\ell m}(\kappa) &= \frac{\tilde{C}_{\ell m}\kappa^3}{2\pi^2} \mathbf{M}'^{(1)}_{\sigma\ell m}(\kappa) \\ \mathbf{B}_{\sigma\ell m}(\kappa) &= \frac{\tilde{C}_{\ell m}\kappa^3}{2\pi^2} \mathbf{N}'^{(1)}_{\sigma\ell m}(\kappa)\end{aligned}, \tag{B-5}$$

which is obtained by using the orthogonal properties of two functions $\mathbf{M}$ and $\mathbf{N}$ as defined in Appendix A. To solve $\mathbf{G}_{m0}$, we first expand it in the following form:

$$\mathbf{G}_{m0}(\mathbf{r},\mathbf{r}') \equiv \frac{4\pi}{c} \int_0^\infty d\kappa \frac{\kappa^3}{2\pi^2} \sum_{\sigma\ell m} h(\kappa) \tilde{C}_{\ell m} \left[ \mathbf{N}_{\sigma\ell m}^{(1)}(\kappa) \mathbf{M}'^{(1)}_{\sigma\ell m}(\kappa) + \mathbf{M}_{\sigma\ell m}^{(1)}(\kappa) \mathbf{N}'^{(1)}_{\sigma\ell m}(\kappa) \right], \tag{B-6}$$

and then substitute it and Eq. (B-4) into Eq. (B-3) to obtain the unknown coefficient $h(\kappa) = \frac{1}{\kappa^2 - k^2}$. With the help of the residue theorem in complex analysis, we have:

$$\int_0^\infty d\kappa \frac{\kappa^2}{\kappa^2 - k^2} j_\ell(\kappa r) j_\ell(\kappa r') = \frac{i\pi k}{2} \left[ h_\ell^{(1)}(kr) j_\ell(kr') H(r-r') + j_\ell(kr) h_\ell^{(1)}(kr') H(r'-r) \right], \tag{B-7}$$

where the Heaviside step function $H(r-r') = \begin{cases} 1, r > r' \\ 0, r < r' \end{cases}$. Recalling $\mathbf{M}_{\sigma\ell m}^{(1)}(\kappa) = \nabla \times (\psi_{\sigma\ell m}^{(1)} \mathbf{r})$ and $\mathbf{N}_{\sigma\ell m}^{(1)}(\kappa) = \frac{1}{\kappa} \nabla \times \nabla \times (\psi_{\sigma\ell m}^{(1)} \mathbf{r})$, we apply Eq. (B-7) and thus $\mathbf{G}_{m0}$ becomes:

$$\mathbf{G}_{m0}(\mathbf{r},\mathbf{r}') = H(r-r') \mathbf{G}_{m0}^{r>r'}(\mathbf{r},\mathbf{r}') + H(r'-r) \mathbf{G}_{m0}^{r<r'}(\mathbf{r},\mathbf{r}'), \tag{B-8}$$

where

$$\begin{cases} \mathbf{G}_{m0}^{r>r'}(\mathbf{r},\mathbf{r}') = \dfrac{ik^2}{c}\sum_{\sigma\ell m}\tilde{C}_{\ell m}\left[\mathbf{N}_{\sigma\ell m}^{(3)}(k)\mathbf{M}_{\sigma\ell m}'^{(1)}(k)+\mathbf{M}_{\sigma\ell m}^{(3)}(k)\mathbf{N}_{\sigma\ell m}'^{(1)}(k)\right] \\ \mathbf{G}_{m0}^{r<r'}(\mathbf{r},\mathbf{r}') = \dfrac{ik^2}{c}\sum_{\sigma\ell m}\tilde{C}_{\ell m}\left[\mathbf{N}_{\sigma\ell m}^{(1)}(k)\mathbf{M}_{\sigma\ell m}'^{(3)}(k)+\mathbf{M}_{\sigma\ell m}^{(1)}(k)\mathbf{N}_{\sigma\ell m}'^{(3)}(k)\right] \end{cases}. \quad \text{(B-9)}$$

Taking the curl of $\mathbf{G}_{m0}$, we obtain:

$$\nabla\times\mathbf{G}_{m0}(\mathbf{r},\mathbf{r}') = H(r-r')\nabla\times\mathbf{G}_{m0}^{r>r'}(\mathbf{r},\mathbf{r}') + H(r'-r)\nabla\times\mathbf{G}_{m0}^{r<r'}(\mathbf{r},\mathbf{r}') + \dfrac{4\pi}{c}(\mathbf{I}-\mathbf{e}_r\mathbf{e}_r)\delta(\mathbf{r}-\mathbf{r}')$$

, (B-10)

where the discontinuity for $\mathbf{G}_{m0}$ at $r=r'$:

$$\mathbf{e}_r\times\left[\mathbf{G}_{m0}^{r>r'}(\mathbf{r},\mathbf{r}')\big|_{r=r'} - \mathbf{G}_{m0}^{r<r'}(\mathbf{r},\mathbf{r}')\big|_{r=r'}\right] = \dfrac{4\pi}{c}(\mathbf{I}-\mathbf{e}_r\mathbf{e}_r)\dfrac{\delta(\theta-\theta')\delta(\varphi-\varphi')}{r^2\sin\theta} \quad \text{(B-11)}$$

is used. Finally, we substitute Eq. (B-10) into Eq. (B-2) to yield the free-space electric DGF $\mathbf{G}_{e0}$ as follows:

$$\mathbf{G}_{e0}(\mathbf{r},\mathbf{r}') = -\dfrac{4\pi}{ck^2}\delta(\mathbf{r}-\mathbf{r}')\mathbf{e}_r\mathbf{e}_r + \dfrac{ik}{c}\sum_{\sigma\ell m}\tilde{C}_{\ell m}\begin{cases}\mathbf{M}_{\sigma\ell m}^{(3)}(k)\mathbf{M}_{\sigma\ell m}'^{(1)}(k)+\mathbf{N}_{\sigma\ell m}^{(3)}(k)\mathbf{N}_{\sigma\ell m}'^{(1)}(k), r>r' \\ \mathbf{M}_{\sigma\ell m}^{(1)}(k)\mathbf{M}_{\sigma\ell m}'^{(3)}(k)+\mathbf{N}_{\sigma\ell m}^{(1)}(k)\mathbf{N}_{\sigma\ell m}'^{(3)}(k), r<r'\end{cases},$$

(B-12)

which is the result in Eq. (7).

## Appendix C: The DGF formulation for a TI sphere

We consider a simple structure ($N=1$ in Fig. 2) which is a non-magnetic TI sphere $(\varepsilon_1,\Theta_1)$ surrounded by a medium $(\varepsilon_2,\Theta_2=0)$ as illustrated in Fig. 12. The source $s$ is in the outer region ($s=2$) and the field $f$ is everywhere ($f=1,2$). Note that the electric DGF has been presented in Eq. (4), including the free-space electric DGF $\mathbf{G}_{e0}$ (Eq. (7)) and the scattering electric DGF $\mathbf{G}_{es}^{(2,2)}$ (Eq. (9)) and $\mathbf{G}_{es}^{(1,2)}$ (Eq. (11)). Using the appropriate boundary conditions of the electric DGFs at $r=R_1$ (Eq. (12)), we obtain the eight simultaneous equations (Eqs. (13)-(20)) with $\alpha_{MM,\ell}^2=\beta_{NN,\ell}^2=\alpha_{NM,\ell}^2=\beta_{MN,\ell}^2=0$ and $\gamma_{MM,\ell}^1=\chi_{NN,\ell}^1=\gamma_{NM,\ell}^1=\chi_{MN,\ell}^1=0$.

Solving both Eq. (16) and Eq. (20), we have:

$$\alpha_{NM,\ell}^1 = \bar{\alpha}_1\tilde{\chi}_1\alpha_{MM,\ell}^1, \quad \text{(C-1)}$$

where $\bar{\alpha}_1 = \dfrac{\alpha}{\pi}\Theta_1$ and $\tilde{\chi}_1 \equiv -i\dfrac{j_\ell(\rho_{1,1})\xi_\ell^{(3)}(\rho_{2,1})}{\sqrt{\varepsilon_2}h_\ell^{(1)}(\rho_{2,1})\xi_\ell^{(1)}(\rho_{1,1})-\sqrt{\varepsilon_1}j_\ell(\rho_{1,1})\xi_\ell^{(3)}(\rho_{2,1})}$.

Substituting Eq. (C-1) into Eq. (17) and solving both Eq. (13) and Eq. (17), we have two coefficients $\gamma_{MM,\ell}^2$ and $\alpha_{MM,\ell}^1$ as follows:

$$\gamma_{MM,\ell}^2 = \frac{-j_\ell(\rho_{2,1})\left[i\bar{\alpha}_1^2\tilde{\chi}_1\xi_\ell^{(1)}(\rho_{1,1}) - \sqrt{\varepsilon_1}\xi_\ell^{(1)}(\rho_{1,1})\right] - j_\ell(\rho_{1,1})\sqrt{\varepsilon_2}\xi_\ell^{(1)}(\rho_{2,1})}{h_\ell^{(1)}(\rho_{2,1})\left[i\bar{\alpha}_1^2\tilde{\chi}_1\xi_\ell^{(1)}(\rho_{1,1}) - \sqrt{\varepsilon_1}\xi_\ell^{(1)}(\rho_{1,1})\right] + j_\ell(\rho_{1,1})\sqrt{\varepsilon_2}\xi_\ell^{(3)}(\rho_{2,1})}, \quad \text{(C-2)}$$

and

$$\alpha_{MM,\ell}^1 = \frac{-h_\ell^{(1)}(\rho_{2,1})\sqrt{\varepsilon_2}\xi_\ell^{(1)}(\rho_{2,1}) + j_\ell(\rho_{2,1})\sqrt{\varepsilon_2}\xi_\ell^{(3)}(\rho_{2,1})}{h_\ell^{(1)}(\rho_{2,1})\left[i\bar{\alpha}_1^2\tilde{\chi}_1\xi_\ell^{(1)}(\rho_{1,1}) - \sqrt{\varepsilon_1}\xi_\ell^{(1)}(\rho_{1,1})\right] + j_\ell(\rho_{1,1})\sqrt{\varepsilon_2}\xi_\ell^{(3)}(\rho_{2,1})}. \quad \text{(C-3)}$$

Substituting Eq. (C-3) into Eq. (C-1), we obtain:

$$\alpha_{NM,\ell}^1 = \bar{\alpha}_1\tilde{\chi}_1 \frac{-h_\ell^{(1)}(\rho_{2,1})\sqrt{\varepsilon_2}\xi_\ell^{(1)}(\rho_{2,1}) + j_\ell(\rho_{2,1})\sqrt{\varepsilon_2}\xi_\ell^{(3)}(\rho_{2,1})}{h_\ell^{(1)}(\rho_{2,1})\left[i\bar{\alpha}_1^2\tilde{\chi}_1\xi_\ell^{(1)}(\rho_{1,1}) - \sqrt{\varepsilon_1}\xi_\ell^{(1)}(\rho_{1,1})\right] + j_\ell(\rho_{1,1})\sqrt{\varepsilon_2}\xi_\ell^{(3)}(\rho_{2,1})}. \quad \text{(C-4)}$$

Next we solve Eq. (14) and Eq. (18) to yield:

$$\beta_{MN,\ell}^1 = \bar{\alpha}_1\tilde{\chi}_2\beta_{NN,\ell}^1, \quad \text{(C-5)}$$

where $\tilde{\chi}_2 \equiv -i\dfrac{h_\ell^{(1)}(\rho_{2,1})\xi_\ell^{(1)}(\rho_{1,1})}{\sqrt{\varepsilon_2}j_\ell(\rho_{1,1})\xi_\ell^{(3)}(\rho_{2,1}) - \sqrt{\varepsilon_1}h_\ell^{(1)}(\rho_{2,1})\xi_\ell^{(1)}(\rho_{1,1})}$. Substituting Eq. (C-5) to

Eq. (19), we solve both Eq. (15) and Eq. (19) to obtain:

$$\chi_{NN,\ell}^2 = \frac{-\xi_\ell^{(1)}(\rho_{2,1})\left[i\bar{\alpha}_1^2\tilde{\chi}_2 j_\ell(\rho_{1,1}) - \sqrt{\varepsilon_1}j_\ell(\rho_{1,1})\right] - \xi_\ell^{(1)}(\rho_{1,1})\sqrt{\varepsilon_2}j_\ell(\rho_{2,1})}{\xi_\ell^{(3)}(\rho_{2,1})\left[i\bar{\alpha}_1^2\tilde{\chi}_2 j_\ell(\rho_{1,1}) - \sqrt{\varepsilon_1}j_\ell(\rho_{1,1})\right] + \xi_\ell^{(1)}(\rho_{1,1})\sqrt{\varepsilon_2}h_\ell^{(1)}(\rho_{2,1})}, \quad \text{(C-6)}$$

and

$$\beta_{NN,\ell}^1 = \frac{-\xi_\ell^{(3)}(\rho_{2,1})\sqrt{\varepsilon_2}j_\ell(\rho_{2,1}) + \xi_\ell^{(1)}(\rho_{2,1})\sqrt{\varepsilon_2}h_\ell^{(1)}(\rho_{2,1})}{\xi_\ell^{(3)}(\rho_{2,1})\left[i\bar{\alpha}_1^2\tilde{\chi}_2 j_\ell(\rho_{1,1}) - \sqrt{\varepsilon_1}j_\ell(\rho_{1,1})\right] + \xi_\ell^{(1)}(\rho_{1,1})\sqrt{\varepsilon_2}h_\ell^{(1)}(\rho_{2,1})}. \quad \text{(C-7)}$$

Substituting Eq. (C-7) into Eq. (C-5), we have:

$$\beta_{MN,\ell}^1 = \bar{\alpha}_1\tilde{\chi}_2 \frac{-\xi_\ell^{(3)}(\rho_{2,1})\sqrt{\varepsilon_2}j_\ell(\rho_{2,1}) + \xi_\ell^{(1)}(\rho_{2,1})\sqrt{\varepsilon_2}h_\ell^{(1)}(\rho_{2,1})}{\xi_\ell^{(3)}(\rho_{2,1})\left[i\bar{\alpha}_1^2\tilde{\chi}_2 j_\ell(\rho_{1,1}) - \sqrt{\varepsilon_1}j_\ell(\rho_{1,1})\right] + \xi_\ell^{(1)}(\rho_{1,1})\sqrt{\varepsilon_2}h_\ell^{(1)}(\rho_{2,1})}. \quad \text{(C-8)}$$

Eq. (14) can be written as follows:

$$\chi_{MN,\ell}^2 = \frac{j_\ell(\rho_{1,1})}{h_\ell^{(1)}(\rho_{2,1})}\beta_{MN,\ell}^1, \quad \text{(C-9)}$$

and we substitute Eq. (C-8) into Eq. (C-9) to obtain:

$$\chi_{MN,\ell}^2 = \frac{\bar{\alpha}_1 \tilde{\chi}_2 j_\ell(\rho_{1,1})}{h_\ell^{(1)}(\rho_{2,1})} \frac{-\xi_\ell^{(3)}(\rho_{2,1})\sqrt{\varepsilon_2} j_\ell(\rho_{2,1}) + \xi_\ell^{(1)}(\rho_{2,1})\sqrt{\varepsilon_2} h_\ell^{(1)}(\rho_{2,1})}{\xi_\ell^{(3)}(\rho_{2,1})\left[i\bar{\alpha}_1^2 \tilde{\chi}_2 j_\ell(\rho_{1,1}) - \sqrt{\varepsilon_1} j_\ell(\rho_{1,1})\right] + \xi_\ell^{(1)}(\rho_{1,1})\sqrt{\varepsilon_2} h_\ell^{(1)}(\rho_{2,1})}$$

. (C-10)

Again, we rewrite Eq. (16) in the following form:

$$\gamma_{NM,\ell}^2 = \frac{\xi_\ell^{(1)}(\rho_{1,1})}{\xi_\ell^{(3)}(\rho_{2,1})} \alpha_{NM,\ell}^1,$$  (C-11)

and we substitute Eq. (C-1) into Eq. (C-11) to obtain:

$$\gamma_{NM,\ell}^2 = \frac{\bar{\alpha}_1 \tilde{\chi}_1 \xi_\ell^{(1)}(\rho_{1,1})}{\xi_\ell^{(3)}(\rho_{2,1})} \frac{-h_\ell^{(1)}(\rho_{2,1})\sqrt{\varepsilon_2}\xi_\ell^{(1)}(\rho_{2,1}) + j_\ell(\rho_{2,1})\sqrt{\varepsilon_2}\xi_\ell^{(3)}(\rho_{2,1})}{h_\ell^{(1)}(\rho_{2,1})\left[i\bar{\alpha}_1^2 \tilde{\chi}_1 \xi_\ell^{(1)}(\rho_{1,1}) - \sqrt{\varepsilon_1}\xi_\ell^{(1)}(\rho_{1,1})\right] + j_\ell(\rho_{1,1})\sqrt{\varepsilon_2}\xi_\ell^{(3)}(\rho_{2,1})}.$$

(C-12)

Hence we obtain the analytical forms of all expansion coefficients $\gamma_{MM,\ell}^2, \chi_{NN,\ell}^2, \gamma_{NM,\ell}^2$, $\chi_{MN,\ell}^2, \alpha_{MM,\ell}^1, \beta_{NN,\ell}^1, \alpha_{NM,\ell}^1$ and $\beta_{MN,\ell}^1$ in Eq. (9) and Eq. (11) for the case of a TI sphere.

**Appendix D**: The expressions of three matrices $\mathbf{T}_\ell^i$, $\mathbf{A}_\ell^i$ and $\mathbf{C}_\ell^i$

We present the forward transfer matrix $\mathbf{T}_\ell^i$ as the form $\mathbf{T}_\ell^i \equiv (\mathbf{A}_\ell^i)^{-1} \mathbf{B}_\ell^i$. The matrix $\mathbf{A}_\ell^i$ is:

$$\mathbf{A}_\ell^i = \begin{pmatrix} A_{11} & 0 & A_{13} & 0 & 0 & 0 & 0 & 0 \\ 0 & 0 & 0 & 0 & 0 & A_{26} & 0 & A_{28} \\ 0 & A_{32} & 0 & A_{34} & 0 & 0 & 0 & 0 \\ 0 & 0 & 0 & 0 & A_{45} & 0 & A_{47} & 0 \\ A_{51} & 0 & A_{53} & 0 & 0 & 0 & 0 & 0 \\ 0 & 0 & 0 & 0 & 0 & A_{66} & 0 & A_{68} \\ 0 & A_{72} & 0 & A_{74} & 0 & 0 & 0 & 0 \\ 0 & 0 & 0 & 0 & A_{85} & 0 & A_{87} & 0 \end{pmatrix}$$  (D-1)

with $A_{11} = j_\ell(\rho_{i+1,i})$, $A_{13} = h_\ell^{(1)}(\rho_{i+1,i})$, $A_{26} = j_\ell(\rho_{i+1,i})$, $A_{28} = h_\ell^{(1)}(\rho_{i+1,i})$, $A_{32} = \xi_\ell^{(1)}(\rho_{i+1,i})$, $A_{34} = \xi_\ell^{(3)}(\rho_{i+1,i})$, $A_{45} = \xi_\ell^{(1)}(\rho_{i+1,i})$, $A_{47} = \xi_\ell^{(3)}(\rho_{i+1,i})$,

$$A_{51} = \frac{ck_{i+1}}{i\omega} \xi_\ell^{(1)}(\rho_{i+1,i}) \quad , \quad A_{53} = \frac{ck_{i+1}}{i\omega} \xi_\ell^{(3)}(\rho_{i+1,i}) \quad , \quad A_{66} = \frac{ck_{i+1}}{i\omega} \xi_\ell^{(1)}(\rho_{i+1,i}) \quad ,$$

$$A_{68} = \frac{ck_{i+1}}{i\omega} \xi_\ell^{(3)}(\rho_{i+1,i}) \quad , \quad A_{72} = \frac{ck_{i+1}}{i\omega} j_\ell(\rho_{i+1,i}) \quad , \quad A_{74} = \frac{ck_{i+1}}{i\omega} h_\ell^{(1)}(\rho_{i+1,i}) \quad ,$$

$$A_{85} = \frac{ck_{i+1}}{i\omega} j_\ell(\rho_{i+1,i}) \quad \text{and} \quad A_{87} = \frac{ck_{i+1}}{i\omega} h_\ell^{(1)}(\rho_{i+1,i}).$$

The matrix $\mathbf{B}_\ell^i$ is:

$$\mathbf{B}_\ell^i = \begin{pmatrix} B_{11} & 0 & B_{13} & 0 & 0 & 0 & 0 & 0 \\ 0 & 0 & 0 & 0 & 0 & B_{26} & 0 & B_{28} \\ 0 & B_{32} & 0 & B_{34} & 0 & 0 & 0 & 0 \\ 0 & 0 & 0 & 0 & B_{45} & 0 & B_{47} & 0 \\ B_{51} & 0 & B_{53} & 0 & B_{55} & 0 & B_{57} & 0 \\ 0 & B_{62} & 0 & B_{64} & 0 & B_{66} & 0 & B_{68} \\ 0 & B_{72} & 0 & B_{74} & 0 & B_{76} & 0 & B_{78} \\ B_{81} & 0 & B_{83} & 0 & B_{85} & 0 & B_{87} & 0 \end{pmatrix} \tag{D-2}$$

with $B_{11} = j_\ell(\rho_{i,i})$, $B_{13} = h_\ell^{(1)}(\rho_{i,i})$, $B_{26} = j_\ell(\rho_{i,i})$, $B_{28} = h_\ell^{(1)}(\rho_{i,i})$, $B_{32} = \xi_\ell^{(1)}(\rho_{i,i})$,

$B_{34} = \xi_\ell^{(3)}(\rho_{i,i})$ , $B_{45} = \xi_\ell^{(1)}(\rho_{i,i})$ , $B_{47} = \xi_\ell^{(3)}(\rho_{i,i})$ , $B_{51} = \frac{ck_i}{i\omega} \xi_\ell^{(1)}(\rho_{i,i})$ ,

$B_{53} = \frac{ck_i}{i\omega} \xi_\ell^{(3)}(\rho_{i,i})$ , $B_{55} = \frac{\alpha}{\pi}(\Theta_{i+1} - \Theta_i)\xi_\ell^{(1)}(\rho_{i,i})$ , $B_{57} = \frac{\alpha}{\pi}(\Theta_{i+1} - \Theta_i)\xi_\ell^{(3)}(\rho_{i,i})$ ,

$B_{62} = \frac{\alpha}{\pi}(\Theta_{i+1} - \Theta_i)\xi_\ell^{(1)}(\rho_{i,i})$ , $B_{64} = \frac{\alpha}{\pi}(\Theta_{i+1} - \Theta_i)\xi_\ell^{(3)}(\rho_{i,i})$ , $B_{66} = \frac{ck_i}{i\omega}\xi_\ell^{(1)}(\rho_{i,i})$ ,

$B_{68} = \frac{ck_i}{i\omega}\xi_\ell^{(3)}(\rho_{i,i})$, $B_{72} = \frac{ck_i}{i\omega} j_\ell(\rho_{i,i})$, $B_{74} = \frac{ck_i}{i\omega} h_\ell^{(1)}(\rho_{i,i})$, $B_{76} = \frac{\alpha}{\pi}(\Theta_{i+1} - \Theta_i) j_\ell(\rho_{i,i})$,

$B_{78} = \frac{\alpha}{\pi}(\Theta_{i+1} - \Theta_i) h_\ell^{(1)}(\rho_{i,i})$, $B_{81} = \frac{\alpha}{\pi}(\Theta_{i+1} - \Theta_i) j_\ell(\rho_{i,i})$, $B_{83} = \frac{\alpha}{\pi}(\Theta_{i+1} - \Theta_i) h_\ell^{(1)}(\rho_{i,i})$,

$B_{85} = \frac{ck_i}{i\omega} j_\ell(\rho_{i,i})$ and $B_{87} = \frac{ck_i}{i\omega} h_\ell^{(1)}(\rho_{i,i})$.

The matrix $\mathbf{C}_\ell^i$ is:

$$\mathbf{C}_\ell^i = \begin{pmatrix} C_1 \\ 0 \\ C_3 \\ 0 \\ C_5 \\ 0 \\ C_7 \\ 0 \end{pmatrix} \tag{D-3}$$

with $C_1 = -j_\ell(\rho_{i+1,i})\delta_{i,N}$, $C_3 = -\xi_\ell^{(1)}(\rho_{i+1,i})\delta_{i,N}$, $C_5 = -\dfrac{ck_{i+1}}{i\omega}\xi_\ell^{(1)}(\rho_{i+1,i})\delta_{i,N}$ and

$C_7 = -\dfrac{ck_{i+1}}{i\omega}j_\ell(\rho_{i+1,i})\delta_{i,N}$, where $\delta_{i,N}$ denotes the Kronecker delta function with the property $\delta_{i,N} = \begin{cases} 1, i = N \\ 0, i \neq N \end{cases}$.

**Reference**

[1] Peccei R D and Quinn H R 1977 Phys. Rev. Lett. **38** 1440

[2] Weinberg S 1978 Phys. Rev. Lett. **40** 223

[3] Wilczek F 1978 Phys. Rev. Lett. **40** 279

[4] Preskill J, Wise M B and Wilczek F 1983 Phys. Lett. B **120** 127

[5] Dine M and Fischler W 1983 Phys. Lett. B **120** 137

[6] Duffy L D and Bibber K 2009 New. J. Phys. **11** 105008

[7] Sikivie P 2010 Int. J. Mod. Phys. A **25** 554

[8] Dror J A, Murayama H and Rodd N L 2021 Phys. Rev. D **103** 115004

[9] Sikivie P 1983 Phys. Rev. Lett. **51** 1415

[10] Wilczek F 1987 Phys. Rev. Lett. **58** 1799

[11] Bernevig B A, Hughes T L and Zhang S C 2006 Science **314** 1757

[12] Fu L and Kane C L 2007 Phys. Rev. B **76** 045302

[13] Qi X L, Hughes T L and Zhang S C 2008 Phys. Rev. B **78** 195424

[14] Hasan M Z and Kane C L 2010 Rev. Mod. Phys. **82** 3045

[15] Moore J E 2010 Nature **464** 194

[16] Qi X L and Zhang S C 2011 Rev. Mod. Phys. **83** 1057

[17] Zhang H, Liu C X, Qi X L, Dai X, Fang Z and Zhang S C 2009 Nat. Phys. **5** 438

[18] Liu C X, Qi X L, Zhang H J, Dai X, Fang Z and Zhang S C 2010 Phys. Rev. B **82** 045122

[19] Qi X L, Li R, Zang J and Zhang S C 2009 Science **323** 1184

[20] Ge L, Zhan T, Han D, Liu X and Zi J 2014 Optics Express **22** 30833

[21] Xie H Y, Chang R and Leung P T 2021 Results Phys. **23** 104014. Note that Eq. (13) in this paper should read $\Delta\hat{\omega} = -\dfrac{3c^3}{4\omega^3}\text{Re}(G)$.

[22] Litvinov V 2020 Magnetism in topological insulator (Springer)

[23] Okada K N, Takahashi Y, Mogi M, Yoshimi R, Tsukazaki A, Takahashi K S, Ogawa N, Kawasaki M and Tokura Y 2016 Nat. Commun. **7** 12245

[24] Wu L, Salehi M, Koirala N, Moon J, Oh S and Armitage N P 2016 Science **354** 1124

[25] Dziom V, Shuvaev A, Pimenov A, Astakhov G V, Ames C, Bendias K, Böttcher J, Tkachov G, Hankiewicz E M, Brüne C, Buhmann H and Molenkamp L W 2017 Nat. Commun. **8** 15197

[26] Ge L, Han D and Zi J 2015 Opt. Comm. **354** 225

[27] Chang R, Xie H Y, Wang Y C, Chiang H P and Leung P T 2019 Results Phys. **15** 102744. Note that the statement on page 3 of this paper "$\mathbf{D} = \varepsilon(\mathbf{r})\mathbf{E} + \dfrac{\alpha\Theta}{\pi}\mathbf{B}$

and , respectively (note that the $P_3$ field in [11]…" should read
"$\mathbf{D} = \varepsilon(\mathbf{r})\mathbf{E} + \frac{\alpha\Theta}{\pi}\mathbf{B}$ and $\mathbf{H} = \mathbf{B} - \frac{\alpha\Theta}{\pi}\mathbf{E}$, respectively (note that the $P_3$ field in [11]…".

[28] Gao D, Ye H and Gao L 2022 Optics Express **30** 8399

[29] Duffy D G 2015 Green's Functions with Applications, 2nd ed. (CRC Press).

[30] Martín-Ruiz A, Cambiaso M and Urrutia L F 2015 Phys. Rev. D **92** 125105

[31] Martín-Ruiz A, Cambiaso M and Urrutia L F 2016 Phys. Rev. D **93** 045022

[32] Castaño-Yepes J D, Franca O J, Ramirez-Gutierrez C F and Valle J C D 2020 Physica E **123** 114202

[33] Crosse J A, Fuchs S and Buhmann S Y 2015 Phys. Rev. A **92** 063831

[34] Song G, Xu J P and Yang Y P 2014 EPL **105** 64001

[35] Zeng R, Zhang M, Wang C, Qian X, Li H, Li Q, Yang Y and Zhu S 2019 J. Opt. Soc. Am. B **36** 1890

[36] Rider M S, Buendía A, Abujetas D R, Huidobro P A, Sánchez-Gil J A and Giannini V 2022 ACS Photonics **9** 1483

[37] Raether H 1988 Surface Plasmons on Smooth and Rough Surfaces and on Gratings

[38] Zayats A V, Smolyaninov I I and Maradudin A A, 2005 Physics Reports **408** 131

[39] Wokaun A, Gordon J P and Liao P F 1982 Phys. Rev. Lett. **48** 957

[40] Nie S M and Emery S R 1997 Science **275** 1102

[41] Prodan E, Radloff C, Halas N J and Nordlander P 2003 Science **302** 419

[42] Yin J, Krishnamoorthy H NS, Adamo G, Dubrovkin A M, Chong Y, Zheludev N I and Cesare Soci C 2017 NPG Asia Materials **9** e425

[43] Sekine A and Nomura K 2021 J. Appl. Phys. **129** 141101

[44] Nenno D M, Garcia C A C, Gooth J, Felser C and Narang P 2020 Nat. Rev. Phys. **2** 682

[45] Xie H Y and Leung P T 2020 J. Phys. Commun. **4** 095014

[46] Tai C T 1993 Dyadic Green Functions in Electromagnetic Theory, 2nd ed. (IEEE, New York)

[47] Chew W C 1990 Wave and fields in inhomogeneous media (Wiley-IEEE Press)

[48] Raad S H, Atlasbaf Z, Rodríguez C J Z, Shahabadi M and Mohassel J R 2020 J. Quant. Spectrosc. Radiat. Transf. **256** 107251

[49] Ruppin R 1975 Phys. Rev. B **11** 2871

[50] Moroz A 2005 Ann. Phys. **315** 352

[51] Chance R R, Prock A and Silbey R 1978 Adv. Chem. Phys. **37** 1

[52] Xie H Y, Leung P T and Tsai D P 2009 Solid State Commun. **149** 625

[53] Grushin A G and Cortijo A 2011 Phys. Rev. Lett. **106** 020403

[54] Chang R, Xie H Y and Leung P T 2022 J. Opt. Soc. Am. B **39** 452

[55] Reijnders A A, Tian Y, Sandilands L J, Pohl G, Kivlichan I D, Zhao S Y F, Jia S, Charles M E, Cava R J, Alidoust N, Xu S, Neupane M, Hasan M Z, Wang X, Cheong S W and Burch K S 2014 Phys. Rev. B **89** 075138

[56] Melo T M, Viana D R, Moura-Melo W A, Fonseca J M, Pereira A R 2016 Phys, Lett. A **380** 973

**Figure caption**

[1] An illustration of two regions over which the axion electrodynamics theory is defined.

[2] Configuration of a TI stratified sphere.

[3] (a) An illustration of a metal-coated TI sphere. Normalized decay rate of a (b) vertical and (c) horizontal emitting dipole near the shell of small dimensions $R_1 = 8nm$ and $R_2 = 10nm$ as a function of the normalized emission frequency $\frac{\omega}{\omega_p}$. Note that the case of a non-TI core is shown by the green (black) line which is obtained in the DGF formulation (the electrostatics approximation).

[4] Normalized decay rate as a function of the normalized emission frequency $\frac{\omega}{\omega_p}$ for a (a) vertical and (b) horizontal emitting dipole near the shell of larger dimensions $R_1 = 100nm$ and $R_2 = 120nm$. Note that the case of a non-TI core is shown by the black line, with the inset identifying the different multipolar order resonances, which are shown in (c) by the dash-dotted line only for the vertical dipole.

[5] Comparison with contributions to the normalized decay rate from different multipolar resonance with the order (a) $\ell = 5$, (b) $\ell = 10$, (c) $\ell = 15$, and (d) $\ell = 20$ as a function of the normalized emission frequency $\frac{\omega}{\omega_p}$ only for the vertical dipole. Note that the case of a non-TI core is shown by the black line.

[6] (a) An illustration of a Ag-TI-Ag-coated TI sphere. Normalized decay rate of a (b) vertical and (c) horizontal emitting dipole near the shell of dimensions $R_1 = 50nm$, $R_2 = 60nm$, $R_3 = 110nm$ and $R_4 = 120nm$ as a function of the normalized emission frequency $\frac{\omega}{\omega_p}$. Note that the case of a non-TI core ($\Theta_1 = 0$) and a non-TI shell ($\Theta_3 = 0$) is shown by the black line, with the inset identifying the different multipolar order resonances, which are shown in (d) by the dash-dotted line only for the vertical dipole.

[7] Comparison with contributions to the normalized decay rate from different multipolar resonance with the order (a) $\ell = 5$, (b) $\ell = 10$, (c) $\ell = 20$, and (d) $\ell = 40$ as a function of the normalized emission frequency $\frac{\omega}{\omega_p}$ only for the vertical dipole. Note that the case of a non-TI core ($\Theta_1 = 0$) and a non-TI shell ($\Theta_3 = 0$) is shown by the black line.

[8] (a) An illustration of an alternating metal-TI stratified sphere. Normalized decay rate of a (b) vertical and (c) horizontal emitting dipole near the shell of

dimensions $R_k = 10k, k = 1, 2, ..., 10$ (unit: nm) as a function of the normalized emission frequency $\frac{\omega}{\omega_p}$. Note that the case of a non-TI core ($\Theta_1 = 0$) and non-TI shells ($\Theta_{2k+1} = 0, k = 1, 2, 3, 4$) is shown by the black line, with the inset identifying the different multipolar order resonances, which are shown in (d) by the dash-dotted line only for the vertical dipole.

[9] Comparison with contributions to the normalized decay rate from different multipolar resonance with the order (a) $\ell = 3$, (b) $\ell = 10$, (c) $\ell = 20$, and (d) $\ell = 30$ as a function of the normalized emission frequency $\frac{\omega}{\omega_p}$ only for the vertical dipole. Note that the case of a non-TI core ($\Theta_1 = 0$) and non-TI shells ($\Theta_{2k+1} = 0, k = 1, 2, 3, 4$) is shown by the black line.

[10] (a) Dielectric function of the bulk Bi$_2$Se$_3$ with spin-orbital coupling via the density functional theory [42], as a function of the normalized emission frequency $\frac{\omega}{\omega_p}$. (b) Normalized decay rate as a function of the normalized emission frequency $\frac{\omega}{\omega_p}$ for a vertical emitting dipole near the shell of dimensions $R_1 = 100nm$ and $R_2 = 120nm$. Note that the case of a non-TI core is shown by the black line, with the inset identifying the different multipolar order resonances, which are shown in (c) by the dash-dotted line. Comparison with contributions to the normalized decay rate from different multipolar resonance with the order (d) $\ell = 5$ and (e) $\ell = 10$ as a function of the normalized emission frequency $\frac{\omega}{\omega_p}$.

[11] (a) Dielectric function of the 5QL Bi$_2$Se$_3$ slab with spin-orbital coupling via the density functional theory [42], as a function of the normalized emission frequency $\frac{\omega}{\omega_p}$. (b) Normalized decay rate as a function of the normalized emission frequency $\frac{\omega}{\omega_p}$ for a vertical emitting dipole near the shell of dimensions $R_1 = 5nm$ and $R_2 = 7nm$. Note that the case of a non-TI core is shown by the black line, with the inset identifying the different multipolar order resonances, which are shown in (c) by the dash-dotted line. Comparison with contributions to the normalized decay rate from different multipolar resonance with the order (d) $\ell = 5$ and (e) $\ell = 10$ as a function of the normalized emission frequency $\frac{\omega}{\omega_p}$.

[12] Geometry of a TI sphere.

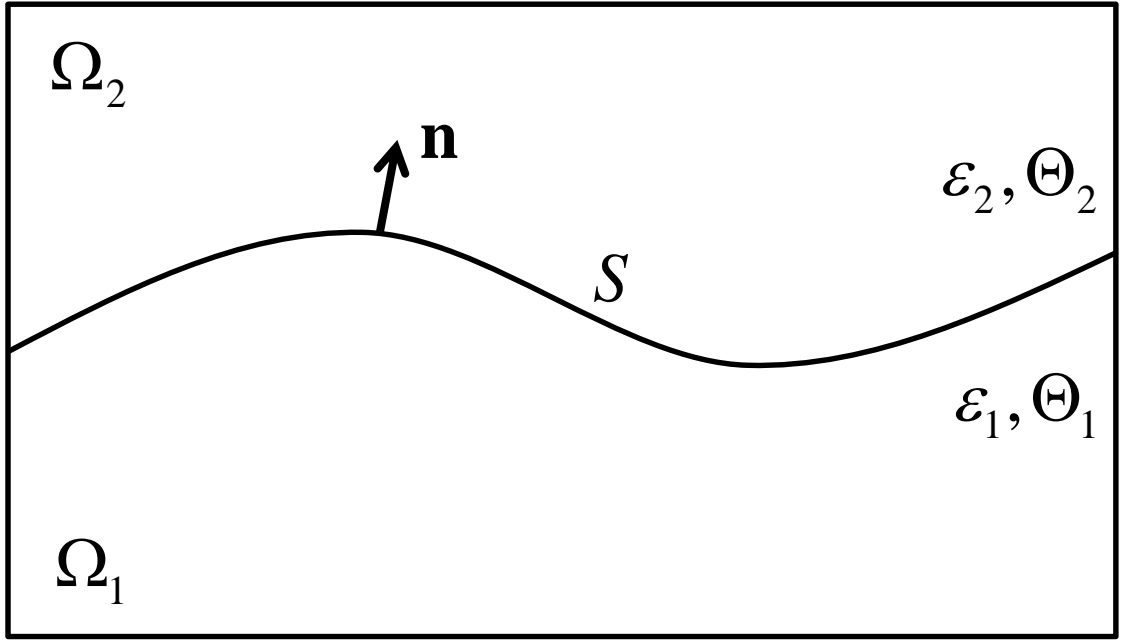

**Fig. 1**

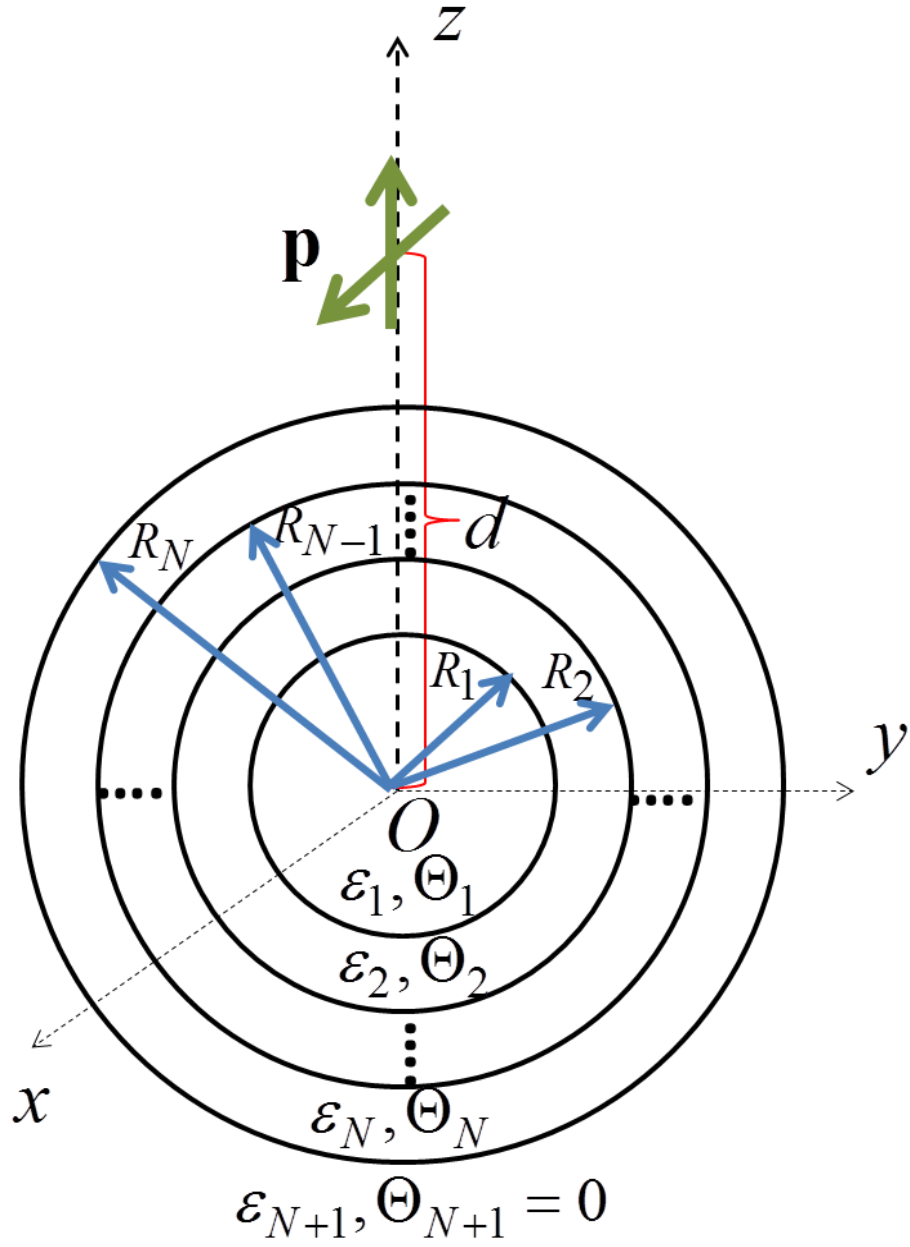

**Fig. 2**

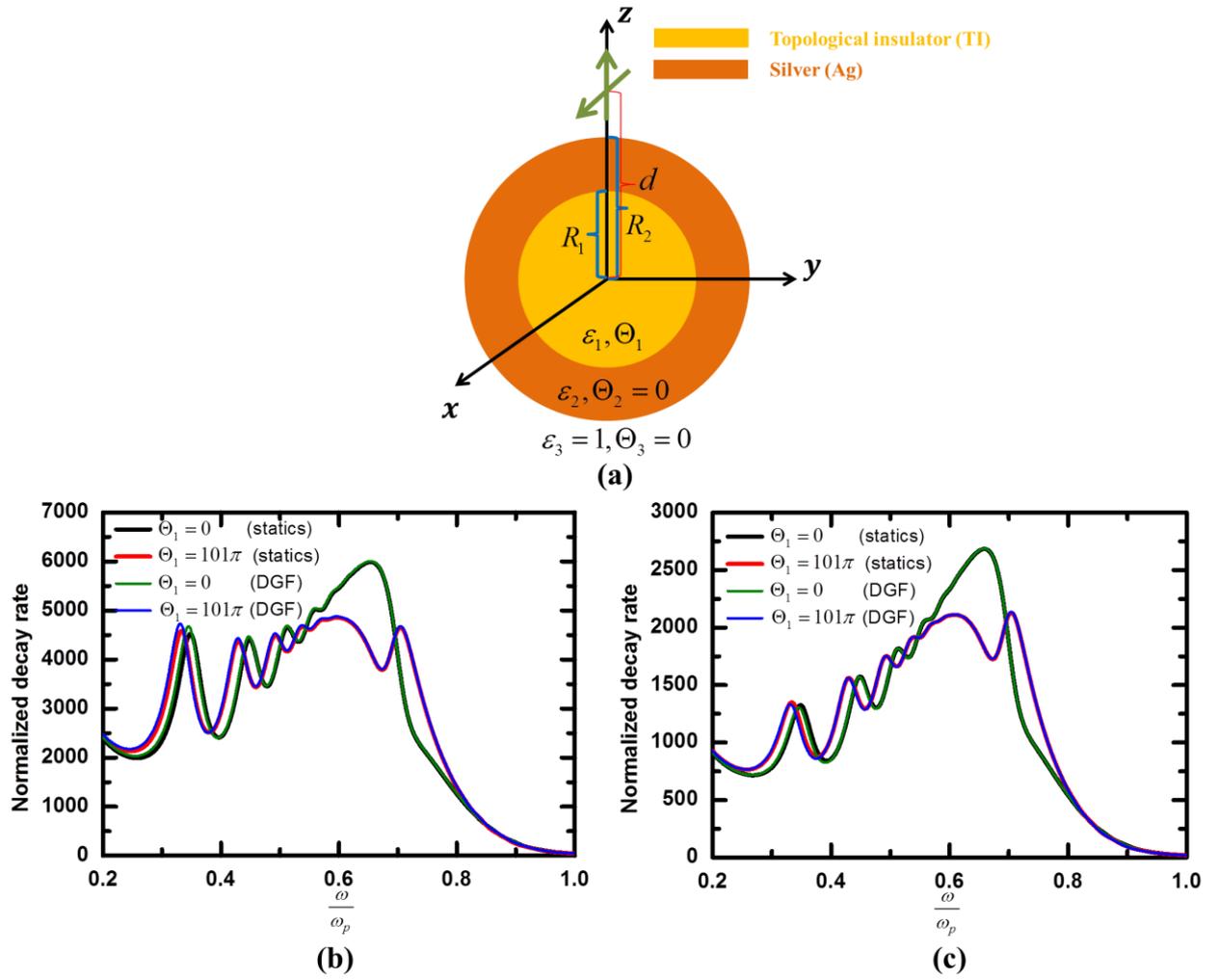

**Fig. 3**

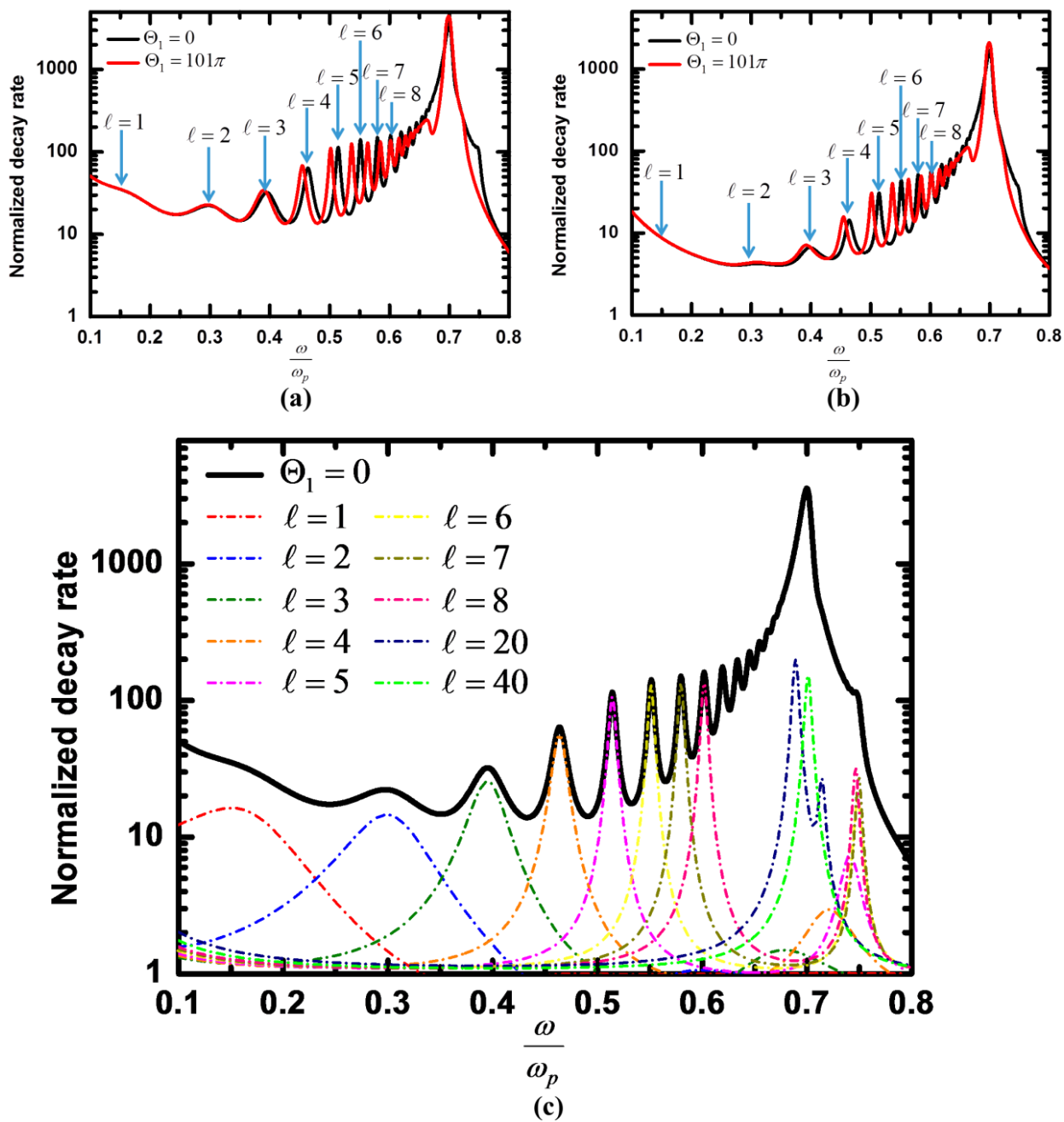

**Fig. 4**

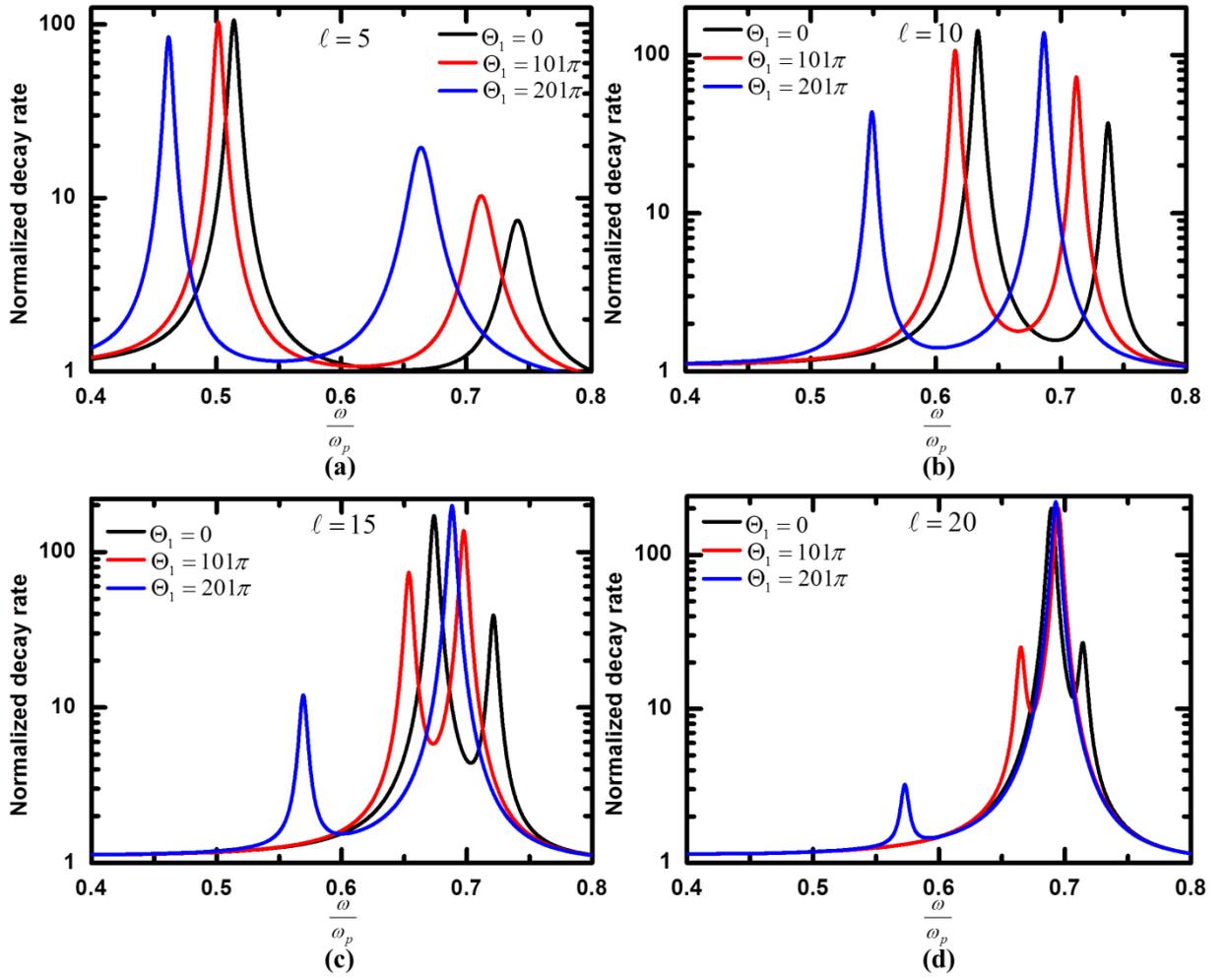

**Fig. 5**

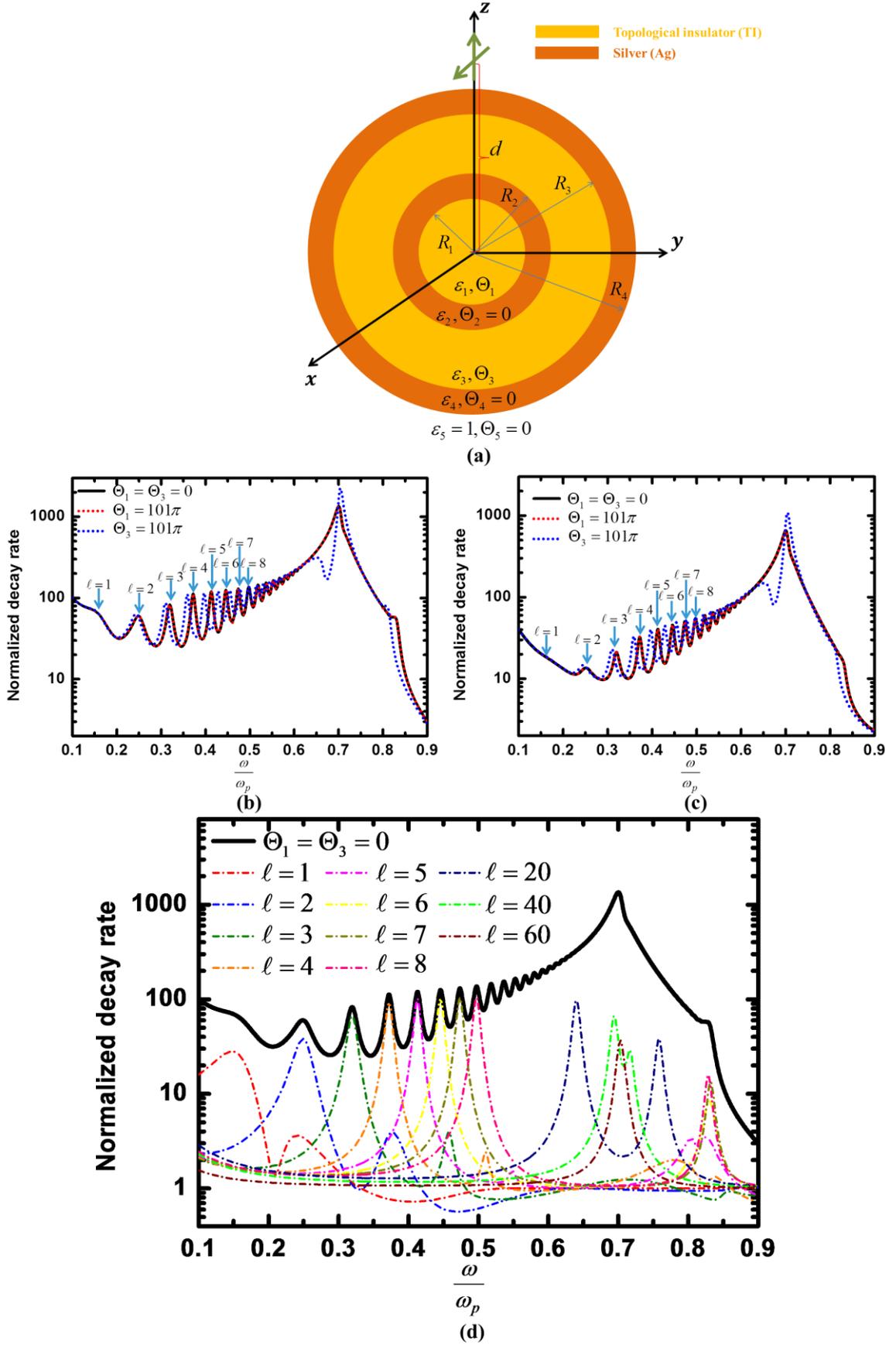

**Fig. 6**

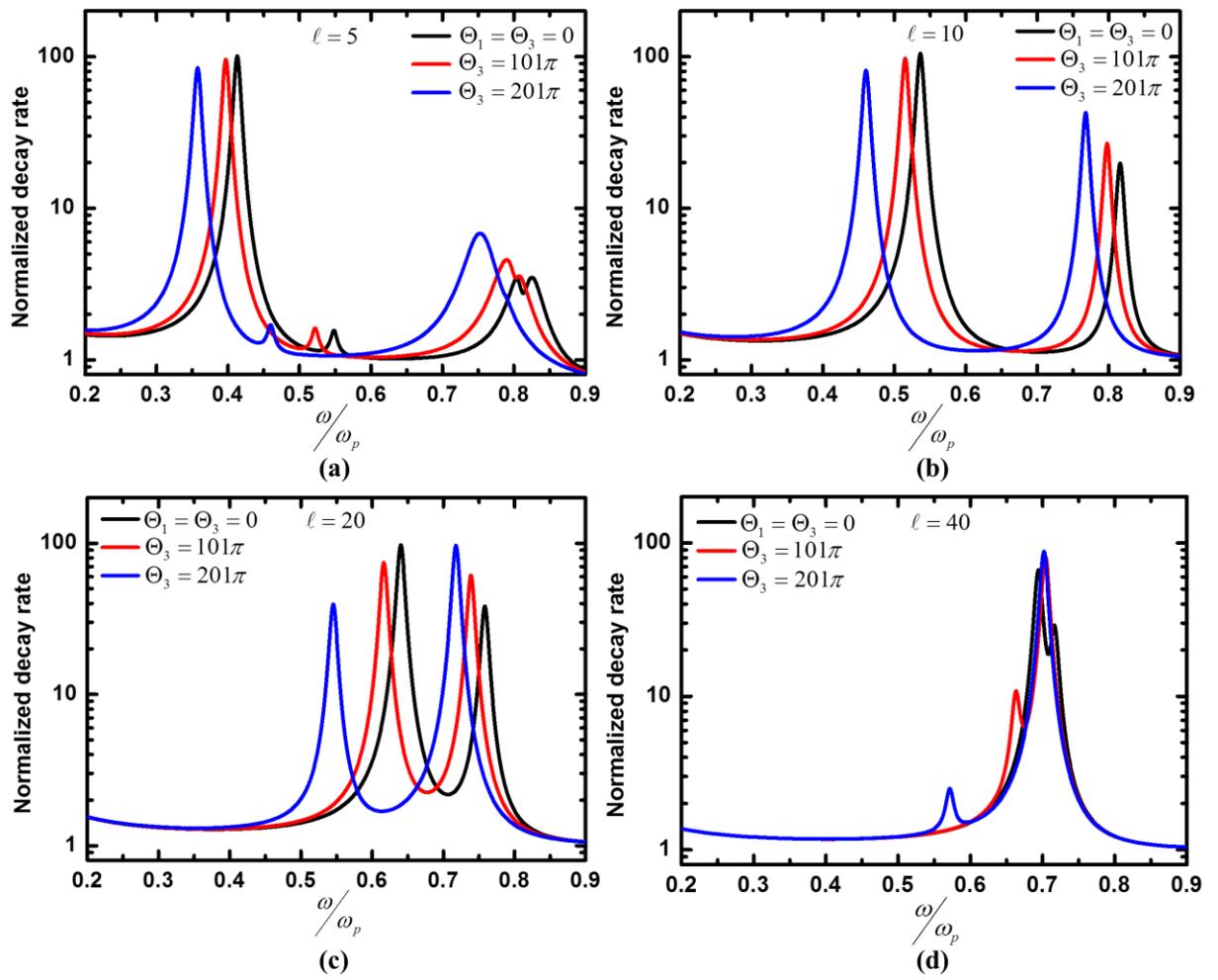

**Fig. 7**

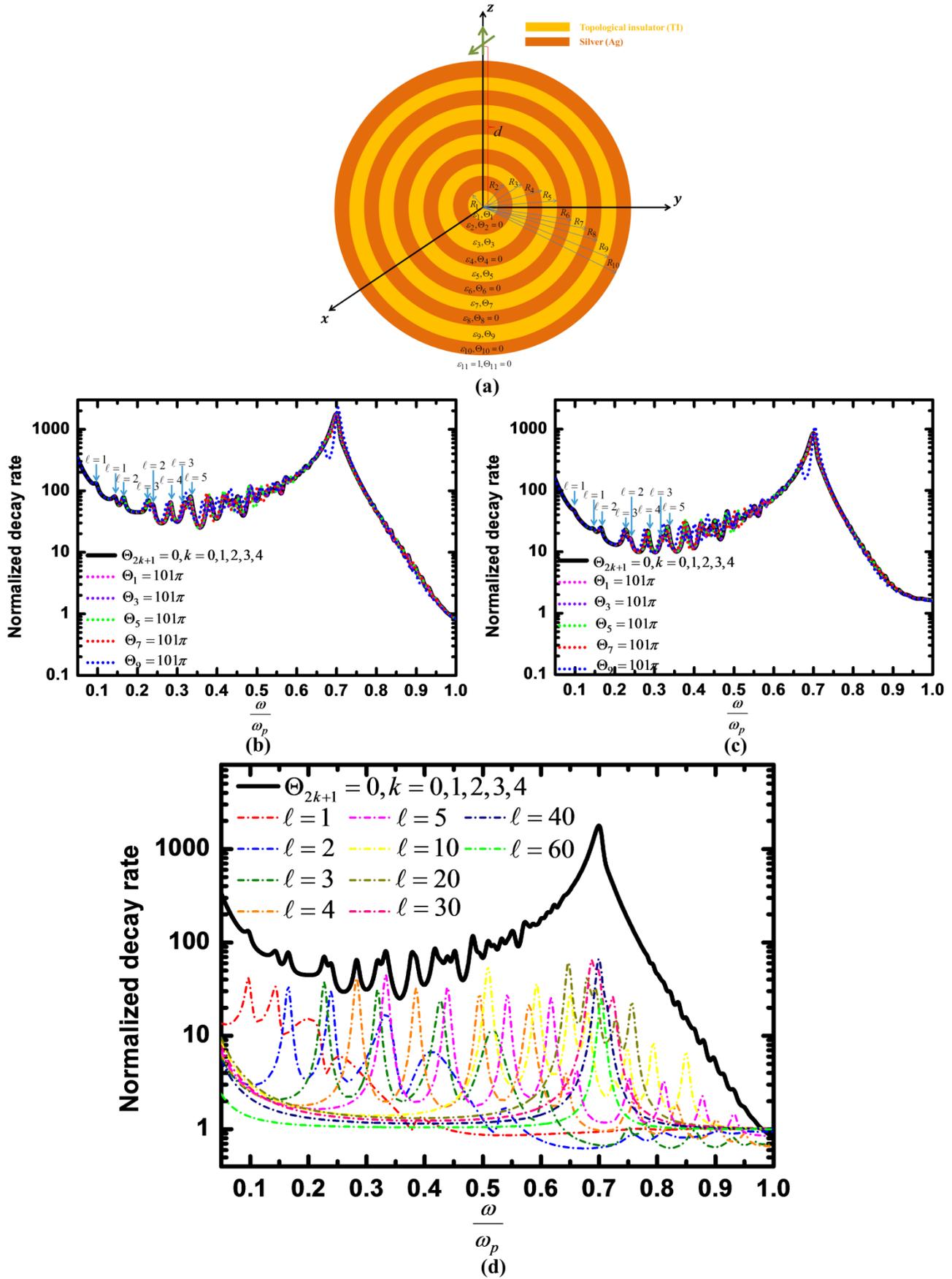

**Fig. 8**

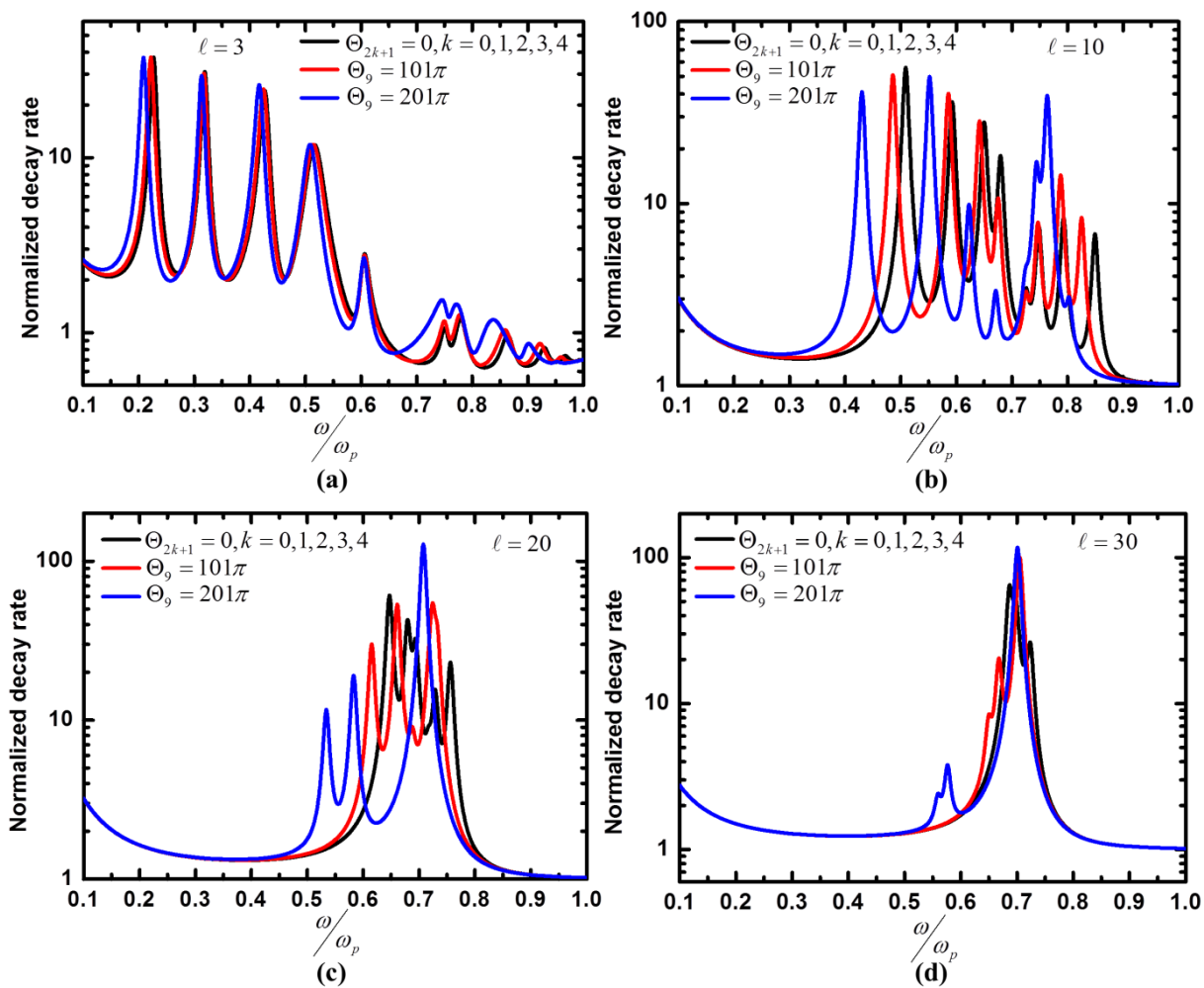

**Fig. 9**

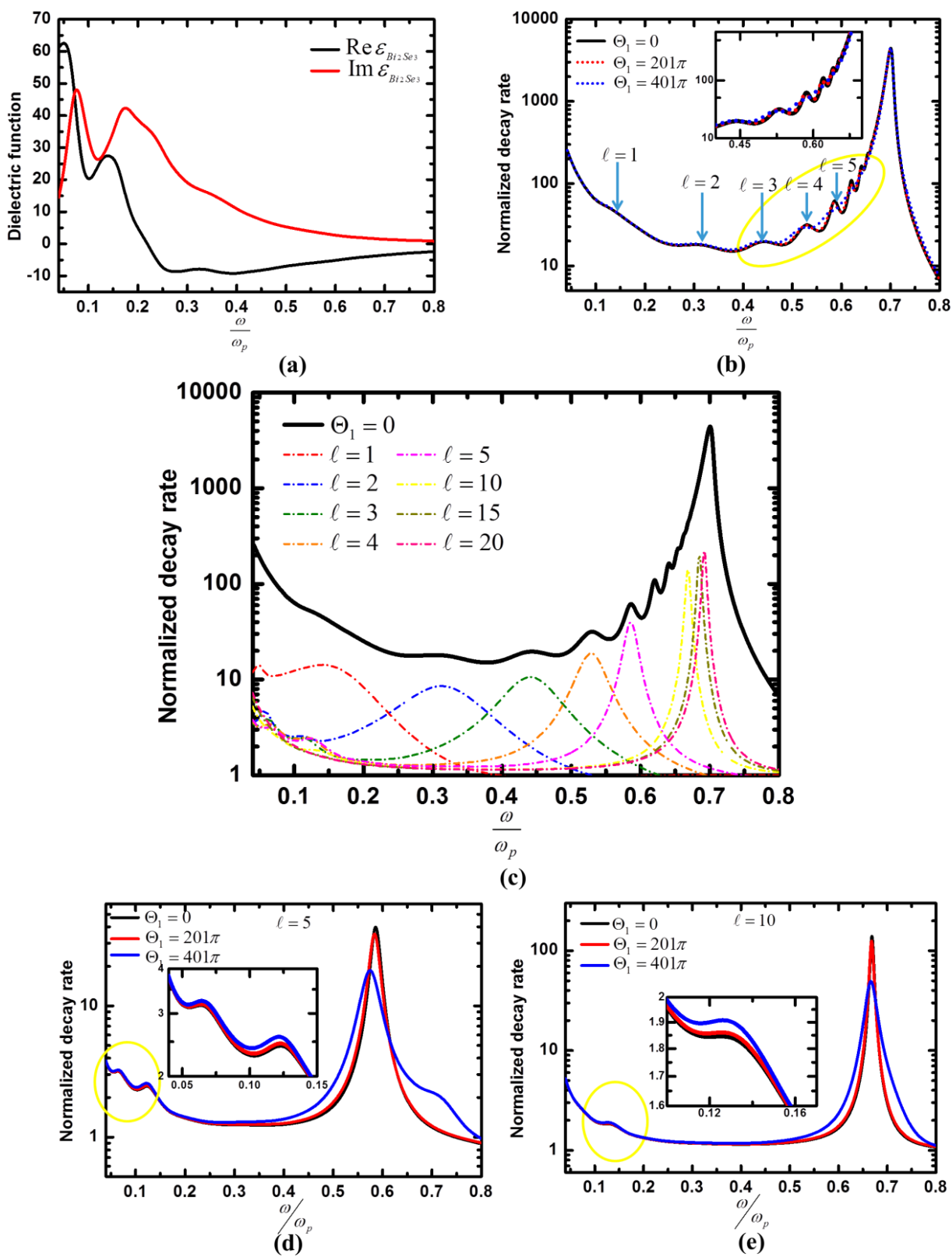

**Fig. 10**

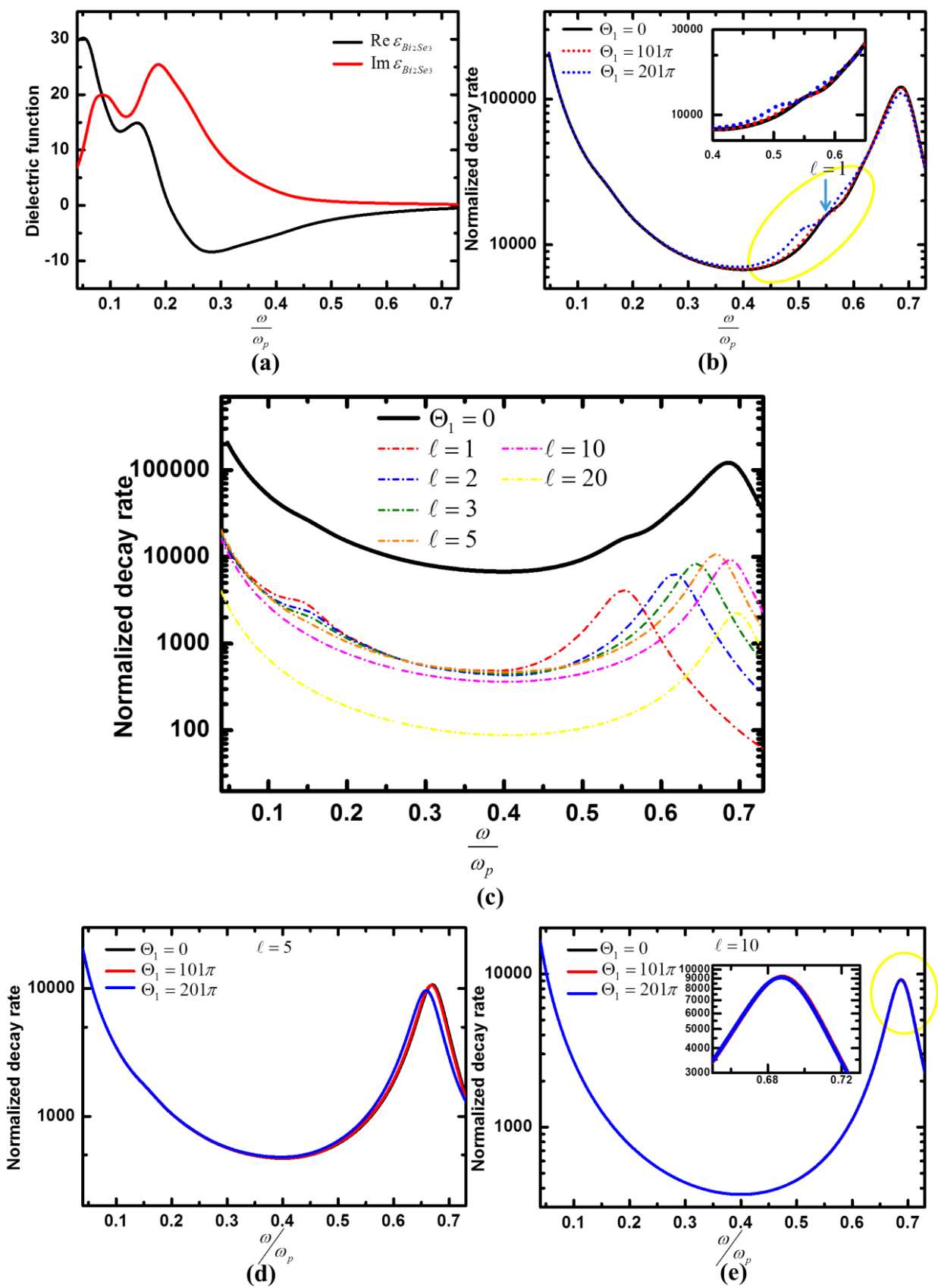

**Fig. 11**

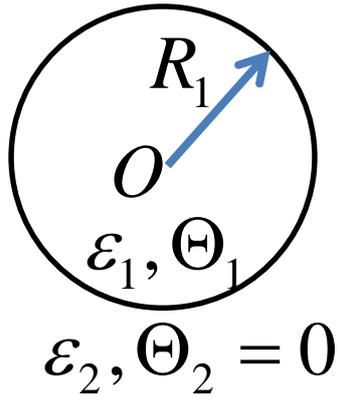

**Fig. 12**